\documentclass[11pt]{article}
\usepackage{amssymb,amsmath}
\usepackage{bm} 
\usepackage{booktabs} 
\usepackage{array}
\usepackage{latexsym}
\usepackage{graphicx}
\usepackage{color}
\usepackage{datetime}
\usepackage[nosort]{cite}
\usepackage{verbatim}
\usepackage{enumerate}
\usepackage{enumitem}
\usepackage{chngpage} 
\usepackage{mathrsfs}
\usepackage{euscript}
\usepackage{psfrag}

\usepackage{mciteplus}

\usepackage[colorlinks=true,      linkcolor=blue,      urlcolor=blue,
            filecolor=blue,      citecolor=blue,       pdfstartview=FitH,
                        pdfpagemode=UseNone,      bookmarksopen=true]{hyperref}  
\usepackage[all]{hypcap}     

\usepackage{tensor}
\usepackage{physics}
\usepackage{tikz}
\usepackage{mathtools}
\usepackage[normalem]{ulem}

\renewcommand{\comm}[1]{} 

\DeclareMathOperator*{\arctanh}{arctanh}

\def\({\left(}
\def\){\right)}
\def\[{\left[}
\def\]{\right]}

\def\coeff#1#2{{\textstyle \frac{#1}{#2}}}

\def\One{{\hbox{ 1\kern-.8mm l}}}

\def\barray{\begin{array}}
\def\earray{\end{array}}
\def\be{\begin{equation}}
\def\ee{\end{equation}}
\def\bea{\begin{eqnarray}}
\def\eea{\end{eqnarray}}
\def\bal{\begin{align}}
\def\eal{\end{align}}

\numberwithin{equation}{section} 

\definecolor{cardinal}{rgb}{0.6,0,0}
\definecolor{darkgreen}{rgb}{0,0.4,0}
\definecolor{golden}{rgb}{0.92, 0.7, 0}
\definecolor{midnight}{rgb}{0, 0, 0.5}
\definecolor{darkblue}{rgb}{0, 0, 0.7}
\definecolor{darkred}{rgb}{0.6, 0, 0}
\definecolor{purple}{rgb}{0.5, 0, 0.5}

\def\Neql#1{{\cal N}\!=\!{#1}}

\def\IR{\mathbb{R}}

\def\IT{\mathbb{T}}

\def\ZZ{\mathbb{Z}}

\def\cB{{\cal B}}

\def\cD{{\cal D}}
\def\cF{{\cal F}}

\def\cI{{\cal I}}

\def\cK{{\cal K}}
\def\cL{{\cal L}}

\def\cP{{\cal P}}
\def\cQ{{\cal Q}}

\def\cU{{\cal U}}

\def\cX{{\cal X}}

\def\scrC{{ \mathscr{C}}}

\def\RAdS{{R_{AdS}}}
\def\RAdSsq{{R^2_{AdS}}}

\def\muscal{{\lambda}}
\def\newmuscal{{\mu}}
\def\param1{{\sigma}} 

\begin{document}

\phantom{AAA}
\vspace{-10mm}

\begin{flushright}
%
%
\end{flushright}

\vspace{1.9cm}

\begin{center}

{\huge {\bf Elliptical and Purely NS Superstrata }}\\
{\huge {\bf \vspace*{.25cm}  }}

\vspace{1cm}

{\large{\bf { Bogdan Ganchev$^{1}$, Anthony Houppe$^{1}$   and  Nicholas P. Warner$^{1,2,3}$}}}

\vspace{1cm}

$^1$Institut de Physique Th\'eorique, \\
Universit\'e Paris Saclay, CEA, CNRS,\\
Orme des Merisiers, Gif sur Yvette, 91191 CEDEX, France \\[12 pt]
\centerline{$^2$Department of Physics and Astronomy}
\centerline{and $^3$Department of Mathematics,}
\centerline{University of Southern California,} 
\centerline{Los Angeles, CA 90089, USA}

\vspace{10mm} 
{\footnotesize\upshape\ttfamily anthony.houppe @ ipht.fr, warner @ usc.edu} \\

\vspace{2.2cm}
 
\textsc{Abstract}

\end{center}

\begin{adjustwidth}{3mm}{3mm} 

We analyze the BPS equations in the ``superstratum sector'' of three-dimensional gauged supergravity.  We obtain multi-parameter supersymmetric solutions that include  elliptical deformations of the supertubes that underlie standard superstrata.  We uplift the three-dimensional solutions to obtain the corresponding six-dimensional geometries.  This yields new families of elliptically-deformed, ambi-bolar hyper-K\"ahler geometries in four dimensions with a non-tri-holomorphic $U(1)$ isometry.  We also find a new family of scaling superstrata whose S-dual lives entirely within the NS-sector of supergravity, and will thus be more amenable to exact analysis using string probes. In all these new superstrata, including the scaling ones, if the momentum charge is non-zero we find that the ellipse stays away from the degeneration locus in which the ellipse becomes flat. 

\vspace{-1.2mm}
\noindent

\end{adjustwidth}

\thispagestyle{empty}
\newpage


\baselineskip=17pt
\parskip=5pt

\setcounter{tocdepth}{2}
\tableofcontents

\baselineskip=15pt
\parskip=3pt

\newpage

\section{Introduction}
\label{sec:Intro}

Dimensional reduction and consistent truncation are extremely powerful tools in the construction of interesting supergravity solutions.  The idea is to reduce the problem to the most relevant degrees of freedom and then solve this simplified system.  The danger is that one may have frozen out degrees of freedom that are essential to the underlying physics. Indeed, one sometimes finds that the reduced system may form singularities in regimes where the complete theory does not. This is, in microcosm, precisely the perspective of microstate geometries and fuzzballs in supergravity and string theory on horizon formation in classical general relativity: black holes are artefacts of neglecting the essential stringy degrees of freedom (whose low-energy coherent excitations are also often describable within supergravity) needed to resolve the singularity and the information paradox \cite{Bena:2022ldq,Bena:2022rna}.

The power and trade-off's inherent in consistent truncations are particularly apparent in  the sphere compactifications that lead to  lower-dimensional gauged supergravity.  The truncation is typically restricted to the lowest-energy, but non-trivial, excitations on the sphere and these are then exactly described by the gauged supergravity, whose solutions can be uplifted to give an exact solution to the higher dimensional supergravity\footnote{This is what one means by a {\it consistent} truncation.}.  The limitations, in capturing only a subset of all modes, are clear, but the ``lowest-energy excitations'' are often the most important for understanding large-scale structure and long-range dynamics. The inherent danger in suppressing the higher modes  is a loss of   fidelity in revealing the detailed structure.  

The power of the consistent truncation lies in the fact that all the non-trivial dynamics in the sphere directions are handled by the machinery of the truncation, and one only has to solve the dynamics of the lower-dimensional gauged supergravity.  Indeed, in holographic field theory there are many simple solutions in four- and five-dimensional gauged supergravity whose ten or eleven-dimensional uplifts are extremely complicated and either took years to build, or their construction remains an open problem. This is especially true of solutions that have rather little symmetry:  the intrinsic construction in higher dimensions   may be completely out of reach.  The  only practical way to obtain  the higher-dimensional solution is then to use the uplift formulae, and even these can be extremely challenging computationally. (See, for example, \cite{Warner:1983du,Godazgar:2014eza}.)

This paper focuses on new results coming from the consistent truncation of IIB supergravity on $\IT^4 \times S^3$ to a gauged supergravity in three-dimensions with an AdS$_3$ vacuum. There is a long history of such consistent truncations  \cite{Cvetic:2000dm,Cvetic:2000zu, Nicolai:2001ac, Nicolai:2003bp, Nicolai:2003ux,Deger:2014ofa,Samtleben:2019zrh}, however, in \cite{Mayerson:2020tcl}, this work  was extended  to include extra tensor multiplets.   These extra degrees of freedom are essential if one is to capture a subset of {\it superstrata} - a particularly important class of microstate geometries \cite{Bena:2015bea,Bena:2016ypk,Tian:2016ucg,Bena:2016agb,Bena:2017xbt,Heidmann:2019zws,Heidmann:2019xrd,Shigemori:2020yuo}.

The price to pay for the consistent truncation is that one is restricted to the $(k=1,m,n)$ family of superstrata, where the $k=1$ signifies the restriction to the lowest modes on the $S^3$.  The advantage, once again, lies in restricting the dynamics to three-dimensions.  The original goal was to simplify the dynamics so that one could find non-BPS microstate geometries, or ``microstrata,'' and this led to the breakthroughs in \cite{Ganchev:2021pgs,Ganchev:2021ewa}.  The three-dimensional BPS equations were also analyzed in \cite{Houppe:2020oqp,Ganchev:2021iwy}, and contrary to our initial expectations, there are new, non-trivial superstrata that can be constructed in three dimensions   \cite{Houppe:2020oqp,Ganchev:2021iwy}.   Perturbation theory had shown that there should be new branches of superstrata in six-dimensions, (see, for example, \cite{Ceplak:2018pws}), but the explicit construction seemed out of reach  because the six-dimensional metric has rather little symmetry.  Once again this plays into the strength of gauged supergravity in low dimensions.

 Our goal in this paper is to make a careful analysis of a specific sector of the three-dimensional BPS equations and to uplift the result to obtain the geometry in six dimensions.  This uplift leads to a new class of elliptically-deformed, ambi-polar, hyper-K\"ahler metrics for the four-dimensional base space that underlies the new superstrata.  These geometries have only a $U(1)$ isometry, and this $U(1)$ is {\it not} tri-holomorphic. We give complete details of the hyper-K\"ahler structure and the K\"ahler potentials for one of the complex structures.  We also obtain the remaining components of the six-dimensional geometry, including the non-trivial fibration vector  obtained from the self-dual magnetic field on this spatial base.  
 
 We show that once this base metric and fibration vector is fixed, then the remaining six-dimensional quantities are determined, as in \cite{Bena:2011dd,Giusto:2013rxa},  by linear equations on this background.  Most importantly, we find that a subset of our new superstrata that can be dualized so as to live in purely the NS sector of IIB supergravity, or string theory.   
 
Standard superstrata are obtained from adding momentum excitations to the D1-D5 system: these branes source RR-fluxes and the momentum carriers are NS5-F1 density waves, which source NS B-fields. One can, of course, take the S-dual of this configuration to the F1-NS5 frame in which the momentum carriers become D5-D1 density modes. The background geometry of the metric and F1-NS5 sources are completely described in the NS sector of the supergravity, or string theory.  However, the momentum carriers, the D5-D1 density modes, live in the RR sector. 

Our new solutions involve  both the  D5-D1 momentum carriers and some  F1-NS5-metric-scalar density modes that create an elliptical deformation of the $S^3$ geometry. We find a branch of this phase space in which the momentum is carried solely by metric-dilaton density modes and RR-fluxes, and, if one takes the S-dual, this deformed superstratum lives entirely within the NS sector.  We show that, like any other superstratum, there is a scaling regime in which the geometry has a long BTZ throat that is then capped off at a parametrically choosable depth.  We expect these new pure-NS superstrata to be important to the efforts in understanding microstructure and fuzzballs using the non-linear sigma-models of the  string world-sheet  \cite{Martinec:2017ztd,Martinec:2019wzw,Martinec:2020gkv}

In Section \ref{sec:3D-Sugr} we review the relevant three-dimensional supergravity and define a further truncation of this system to a superstratum sector.  Section \ref{sec:genBPS}  contains a careful analysis of the BPS equations for the superstratum sector, and  we provide analytic solutions for a multi-parameter family of BPS solutions that were discovered perturbatively in \cite{Houppe:2020oqp}.   In Section \ref{sec:uplift} we uplift the solution to obtain the complete six-dimensional geometries of the new families of superstrata, and we highlight the new  ambi-polar, hyper-K\"ahler geometries that form the  base of the six-dimensional solutions.    We summarize the details of the purely NS superstrata in Section \ref{sec:Example}, and in  Section \ref{sec:AdS2} we show that these superstrata do indeed have scaling regions and that the supertube remains bounded away from the degenerate flattened ellipse.    
We summarize our results and make some final remarks in Section \ref{sec:Conclusions} .  

\section{The three-dimensional supergravity}
\label{sec:3D-Sugr}

Here we summarize the essential features of the underlying three-dimensional supergravity.  More details may be found in  \cite{Houppe:2020oqp,Ganchev:2021pgs}.

\subsection{The field theory content}
\label{sec:fields}

The $(1,m,n)$ superstrata can be described within an $\Neql{4}$ gauged supergravity theory in three dimensions.  This theory has eight supersymmetries and the spectrum consists of a  graviton, four gravitini, $\psi_\mu^A$, two sets of six vector fields, $A_\mu^{IJ}= -A_\mu^{JI}$ and $B_\mu^{IJ}= -B_\mu^{JI}$, $20$ fermions, $\chi^{\dot A r}$, and  $20$ scalars parametrized by the coset: 
\begin{equation}
\frac{G}{H} ~\equiv~ \frac{SO(4,5)}{SO(4) \times SO(5)} \,.
\label{coset1}
\end{equation}
 The gauge group is an $SO(4) \ltimes \IT^6$ subgroup of $G$ and the $\IT^6$ gauge invariance  is typically fixed by setting six of the scalars to zero.   The  associated gauge fields, $B_\mu^{IJ}$, can then be integrated out.  The result is an action for the graviton, the gravitini, the fermions, the  $SO(4)$ gauge fields, $A_\mu^{IJ}$, and $14$ scalars. Capital Latin indices, $I,\,J$, denote the vector representation of $SO(4)$ and we use small Greek letters, $\mu,\,\nu$, for spacetime indices. The scalars can be described by the non-compact generators of a  $GL(4,\IR)$ matrix, ${P_I}^J$, and an $SO(4)$ vector,  $\chi_I$.  The theory then has a remaining gauge symmetry $SO(4) \subset G$ that acts on the $\chi_I$ and on the left of  ${P_I}^J$, along with a composite, local symmetry $SO(4) \subset H$ that acts on the fermions and on the right of ${P_I}^J$. This part of the composite local symmetry is typically fixed by taking $P$ to be symmetric.  Indeed, once all this gauge fixing is done, the scalars are usually parametrized by a manifestly symmetric matrix, $m_{IJ}$, and its inverse, $m^{IJ}$: 
\begin{equation}
m_{IJ}   ~\equiv~   \big(P  \, P^T\big)_{IJ} \,, \qquad m^{IJ}   ~=~  \big ( (P^{-1})^T\,P^{-1}  \big)^{IJ}  \,.
\label{Mdefn}
\end{equation}

The gauge covariant derivatives are defined in terms of the $SO(4)$-dual vectors:
\begin{equation}
{\widetilde A_\mu}{}^{IJ}  ~\equiv~ \coeff{1}{2} \,\epsilon_{IJKL}\,{A_\mu}^{KL} \,,  
\label{dualGFs}
\end{equation}
with minimal couplings 
\begin{equation}
\cD_\mu \, \cX_{I}  ~=~ \partial_\mu \, \cX_{I}   ~-~  2\, g_0\,\widetilde A_\mu{}^{IJ} \, \cX_{J} \,. 
\label{covderiv}
\end{equation}
In particular, one has:
\begin{equation}
\cD_\mu m_{IJ}   ~=~   \partial_\mu m_{IJ}  ~-~   2\, g_0\,\widetilde A_\mu{}^{IK} m_{KJ }~-~   2\, g_0\,\widetilde A_\mu{}^{JK} m_{IK }   \,   \,.
\label{Dmform}
\end{equation}

The field strengths are thus
\begin{equation}
F_{\mu \nu}{}^{IJ}  ~=~ \coeff{1}{2}\, \epsilon_{IJKL} \,  \widetilde F_{\mu \nu}{}^{KL}    ~=~  \partial_{\mu}   A_{\nu}{}^{IJ}   ~-~  \partial_{\nu}   A_{\mu}{}^{IJ}   ~-~ 2 \,  g_0 \, \big(   A_{\mu} {}^{IL} \,\widetilde  A_{\nu} {}^{LJ}  ~-~ A_{\mu} {}^{JL} \,\widetilde  A_{\nu} {}^{LI}\big)  \,.
\label{fieldstrength}
\end{equation}
The gauge coupling has dimensions of inverse length, and is related to the charges of the D1-D5 system via: 
 \begin{equation}
g_0 ~\equiv~ (Q_1 Q_5)^{-\frac{1}{4}} \,.
\label{g0reln}
\end{equation}

We  also define the following combinations of fields:
\begin{equation}
\begin{aligned}
Y_{\mu \, IJ}  ~\equiv~&  \chi_J \,\cD_\mu \chi_I ~-~  \chi_I \,\cD_\mu\chi_J \,, \qquad  \qquad   C_{\mu}^{IJ}  ~\equiv~  g_{\mu \rho} \,  \varepsilon^{\rho \sigma \nu} \, m_{IK} m_{JL}\,F_{\sigma \nu}^{KL}  \,,   \\
\scrC_{\mu}^{IJ}   ~\equiv~ & P^{-1}{}_{I}{}^K \,  P^{-1}{}_{J}{}^L \, C_\mu^{KL}~=~  g_{\mu \rho}  \, \varepsilon^{\rho \sigma \nu} \,    P{}_{I}{}^K \,  P{}_{J}{}^L \,F_{\sigma \nu}^{KL}\,.
\end{aligned}
\label{cs_definition}
\end{equation}
Our  conventions for $\varepsilon$ can be found in Appendix A of \cite{Ganchev:2021iwy}.

\subsection{The bosonic action}
\label{sec:3Daction}

Following \cite{Houppe:2020oqp},  we will use a metric signature\footnote{Section 2.7 of \cite{Houppe:2020oqp} discusses how to convert to the mostly positive signature.} of $(+- -) $. The action may now be written as  \cite{Mayerson:2020tcl,Houppe:2020oqp} (note that the first reference uses different metric conventions): 
\begin{equation}
\begin{aligned}
\cL ~=~ & -\coeff{1}{4} \,e\,R    ~+~ \coeff{1}{8}\,e \, g^{\mu \nu} \, m^{IJ}   \, (\cD_\mu\, \chi_{I})  \, (\cD_\nu\, \chi_{J})  ~+~  \coeff{1}{16}\,e \, g^{\mu \nu} \,  \big( m^{IK} \, \cD_\mu\, m_{KJ}  \big)   \big( m^{JL} \, \cD_\nu\, m_{LI}  \big)  \\
 &-~   \coeff{1}{8}\, e \, g^{\mu \rho}  \, g^{\nu \sigma} \, m_{IK} \,m_{JL}\,  F_{\mu \nu }^{IJ}  \, F_{\rho \sigma }^{KL}  ~-~e\, V   \\
&+~  \coeff{1}{2}\,e  \, \varepsilon^{\mu \nu \rho} \, \Big[  g_0 \,\big(A_\mu{}^{IJ}\, \partial_\nu  \widetilde A_\rho{}^{IJ}  ~+~\coeff{4}{3}\,  g_0 \, A_\mu{}^{IJ} \,  A_\nu{}^{JK}\, A_\rho{}^{KI} \,\big) ~+~  \coeff{1}{8}\,  {Y_\mu}{}^{IJ}  \, F_{\nu \rho}^{IJ} \Big] \,,
\end{aligned}
\label{eq:3Daction}
\end{equation}

The scalar potential is given by:
\begin{equation}
V ~=~  \coeff{1}{4}\, g_0^2   \,  \det\big(m^{IJ}\big) \, \Big [\, 2 \,\big(1- \coeff{1}{4} \,  (\chi_I \chi_I)\big)^2    ~+~ m_{IJ} m_{IJ}  ~+~\coeff{1}{2} \,  m_{IJ} \chi_I \chi_J  ~-~\coeff{1}{2} \,  m_{II}  \,  m_{JJ}\, \Big] \,.
\label{potential1}
\end{equation}

It is frequently convenient to  fix the local $SO(4)$ gauge symmetry by diagonalizing $P$ in terms of four scalar fields, $\muscal_i$:
\begin{equation}
P  ~=~  {\rm diag} \big(\,  e^{\muscal_1} \,, \,  e^{\muscal_2} \,, \,  e^{\muscal_3} \,, \,  e^{\muscal_4} \, \big) \,.
\label{Pdiag}
\end{equation}
The potential then reduces to:
\begin{equation}
\begin{aligned}
V~=~ & \coeff{1}{4}\, g_0^2   \,e^{-2\, (\muscal_1 +\muscal_2+\muscal_3+\muscal_4)} \, \Big [\, 2 \,\big(1- \coeff{1}{4} \,  (\chi_I \chi_I)\big)^2    ~+~ \big( e^{4\, \muscal_1}+e^{4\, \muscal_2}+e^{4\, \muscal_3}+e^{4\, \muscal_4}  \big)  \\
& \qquad\qquad\qquad\qquad \qquad\qquad~+~\coeff{1}{2}\, \big( e^{2\, \muscal_1}\, \chi_1^2 +e^{2\, \muscal_2}\, \chi_2^2 + e^{2\, \muscal_3}\, \chi_3^2+e^{2\, \muscal_4}\, \chi_4^2  \big) \\
& \qquad\qquad\qquad\qquad \qquad\qquad~-~\coeff{1}{2}\, \big( e^{2\, \muscal_1} +e^{2\, \muscal_2} + e^{2\, \muscal_3} + e^{2\, \muscal_4}  \big)^2  \, \Big] 
 \,.
\end{aligned}
\label{potential2}
\end{equation}

The supersymmetry means that there is a superpotential
\begin{equation}
\begin{aligned}
W  ~\equiv~ & \coeff{1}{4} \, g_0 \,  (\det(P))^{-1}  \,   \Big [\, 2 \,\Big(1- \coeff{1}{4} \,  (\chi_A \chi_A)\Big) ~-~ {\rm Tr}\big(P\, P^T\big)  \, \Big] \\
~=~ & \coeff{1}{4} \, g_0 \,e^{-\muscal_1 -\muscal_2-\muscal_3-\muscal_4} \, \Big [\, 2 \,\Big(1- \coeff{1}{4} \,  (\chi_A \chi_A)\Big) ~-~ \Big( e^{2\, \muscal_1}+e^{2\, \muscal_2}+e^{2\, \muscal_3}+e^{2\, \muscal_4}  \Big) \, \Big]   \,,
 \end{aligned}
\label{superpot}
\end{equation}
and that the potential may be written as
\begin{equation}
V~=~ \delta^{ij} \frac{\partial W}{\partial \muscal_i}  \frac{\partial W}{\partial \muscal_j}  ~+~ 2\, m^{IJ} \, \frac{\partial W}{\partial \chi_I} \frac{\partial W}{\partial \chi_J}   ~-~2\, W^2  \,.
\label{potential3}
\end{equation}

The potential has a supersymmetric critical point\footnote{There are also non-supersymmetric flat directions extending from this supersymmetric critical point.} for  $\muscal_j = \chi_I =0$.   At this point $V$ takes the value
\begin{equation}
V_0 ~=~    -  \coeff{1}{2}\, g_0^2   \,.
\label{susypt}
\end{equation}
Setting all the other fields to zero, the Einstein equations give: 
\begin{equation}
R_{\mu \nu} ~=~ - 4 \, V_0 \, g_{\mu \nu} ~=~    2 \, g_0^2 \, g_{\mu \nu}  \,.
\label{susyvac}
\end{equation}
and the supersymmetric vacuum\footnote{One should note that because we are using a metric signature $(+ - - )$ the cosmological constant of AdS is positive, contrary to the more standard and rational choice of signature.} is an AdS$_3$ of radius, $g_0^{-1}$.  We therefore define
\begin{equation}
 \RAdS ~=~ \frac{1}{g_0} \,,
\label{RAdSscale}
\end{equation}
and so we will henceforth use this to set the overall scale of the metric.

\subsection{The metric}
\label{sec:metric}

The original superstrata were parametrized in terms of the usual double null coordinates, $(u,v)$, which are related to the time and circle coordinates, $(t,y)$, via:
\begin{equation}
u ~\equiv~\frac{1}{\sqrt{2}} \, \big( t ~-~y  \big) \,, \qquad  v ~\equiv~\frac{1}{\sqrt{2}} \, \big(t ~+~y )\,, 
\label{uvtyreln}
\end{equation}
where $y$ is periodically identified as 
\begin{equation}
y ~\equiv~ y ~+~ 2 \pi \,R_y\,.
\label{yperiod}
\end{equation}
It is also convenient to compactify the radial coordinate and make the other coordinates scale-free:
\begin{equation}
\xi ~=~\frac{r}{\sqrt{r^2+ a^2}} \,,  \qquad  \tau~=~ \frac{t}{R_y}\,  \,,  \qquad  \psi~=~ \frac{\sqrt{2}\, v }{R_y}\,, 
\label{xidef}
\end{equation}
where $ 0 \le \xi < 1$, $\psi$ inherits the periodicity $\psi \equiv \psi + 2 \pi$ from (\ref{yperiod}) and $a$ is the radius of the supertube locus.

As noted in \cite{Houppe:2020oqp,Ganchev:2021pgs}, the most general three-dimensional metric can then be recast in the form:
\begin{equation}
ds_{3}^{2}  ~=~  \RAdSsq \, \bigg[ \,\Omega_1^{2} \, \bigg(d \tau +   \frac{k}{(1- \xi^{2})} \, d\psi \bigg)^2~-~\,\frac{\Omega_0^{2}}{(1-\xi^{2} )^{2}} \, \big( d \xi^2 ~+~ \xi^2 \, d \psi^2 \big) \, \bigg] \,,
\label{genmet1}
\end{equation}
for three arbitrary functions $\Omega_0$,  $\Omega_1$ and $k$ of the three coordinates, $(\tau ,\xi,\psi)$.  For later convenience, we also define the metric function:
\begin{equation}
\kappa(\xi) ~\equiv~  \frac{k}{(1- \xi^{2})} \,.
\label{kappadefn}
\end{equation}

If one returns to the coordinates $(t,r,v)$, one obtains the more canonical superstratum metric:
\begin{equation}
ds_{3}^{2}  ~=~   \RAdSsq \, \bigg[ \, \frac{\Omega_1^{2}}{R_y^2} \, \bigg(dt  + \frac{\sqrt{2}}{a^2} \, (r^2  + a^2 ) \, k  \, dv  \bigg)^2~-~  \Omega_0^{2}\,\bigg(\frac{dr^2}{r^2 + a^2} ~+~\frac{2}{R_y^2 \, a^4 } \,r^2\,(r^2 + a^2) \, dv^2 \bigg)  \, \bigg]\,.
\label{genmet2}
\end{equation}
If one further sets:
\begin{equation}
\Omega_0 ~=~ \Omega_1 ~=~ 1\,, \qquad  k ~=~ \xi^2    ~=~  \frac{  r^2 }{ (r^2+ a^2) }  \,,
\label{AdSvals}
\end{equation}
then (\ref{genmet2}) becomes the metric of global AdS$_3$:
\begin{equation}
ds_{3}^{2}  ~=~ \RAdSsq \, \bigg[ \,   \bigg(1 + \frac{r^2}{a^2} \bigg)\,  \bigg(\frac{dt}{R_y}\bigg)^2~-~  \frac{dr^2}{r^2 + a^2}  ~-~ \frac{r^2}{a^2}\,  \bigg(\frac{dy}{R_y}\bigg)^2 \, \bigg] \,.
\label{AdSmet}
\end{equation}
%

\subsection{The superstratum truncation}
\label{sub:qball}

We impose the following  symmetries on the fields:
\begin{itemize}
\item[(i)]  Invariance under internal $U(1)$ that rotates the indices $(3,4)$.
\item[(ii)] Translation  invariance under:  $\tau\to \tau -\alpha $ and   $\psi \to \psi - \beta$
\item[(iii)] Reflection invariance under $\psi \to - \psi $, $t \to - t $, accompanied by a discrete internal $SO(4)$ rotation,  $2 \to -2$, $4 \to -4$.
\end{itemize}

For the fields, $\chi_I$, the first and third symmetries imply that
\begin{equation}
 \chi_2 ~=~   \chi_3 ~=~  \chi_4 ~=~ 0 \,,   \label{trunc1}
\end{equation}
and the second symmetry means that the remaining field, $\chi_1$, is only a function of $\xi$.  

Similarly, the scalar matrix must be diagonal, and the entries are only functions of $\xi$.  We therefore have 
\begin{equation}
\chi_1   ~=~  \sqrt{1 - \xi^2} \,   \nu(\xi)  \,, \quad  \chi_2= \chi_3 = \chi_4    ~=~ 0   \,, \qquad P  ~=~  {\rm diag} \big(\,  e^{\newmuscal_1+ \newmuscal_0} \,, \,  e^{\newmuscal_1- \newmuscal_0} \,, \,  e^{\newmuscal_2} \,, \,  e^{\newmuscal_2} \, \big) \,,
\label{scalaransatz}
\end{equation}
where we have introduced the factor of $\sqrt{1 - \xi^2}$ into $\chi_1$ to match the choices made in  \cite{Ganchev:2021pgs,Ganchev:2021iwy}.  Also note that the first symmetry requirement means that the last two eigenvalues of $P$ must be equal.

The first symmetry also reduces the gauge fields to $\tilde A^{12}$ and $\tilde A^{34}$, while remaining symmetries mean that these fields can only depend on $\xi$, and have no $d\xi$ components.  The gauge fields can therefore be reduced to:  
\begin{equation}
\tilde A^{12} ~=~ \frac{1}{g_0} \,\big[\,  \Phi_1(\xi)  \, d\tau ~+~  \Psi_1(\xi)  \, d\psi \, \big]\,, \qquad  \tilde A^{34} ~=~ \frac{1}{g_0} \,\big[\,\Phi_2(\xi)  \, d\tau ~+~  \Psi_2(\xi)  \, d\psi    \, \big] \,,
  \label{gauge_ansatz}
\end{equation}
where we have  introduced explicit factors of $g_0^{-1}$ so as to cancel the $g_0$'s in the minimal coupling and thus render the fields and interactions scale independent. 

Finally, the time-translation and $\psi$-translational invariance means that $\Omega_0$,  $\Omega_1$  and  $k$, can only depend on $\xi$.  

The Ansatz thus involves eleven arbitrary functions of  one variable, $\xi$:
\begin{equation}
{\cal F} ~\equiv~ \big\{\, \nu  \,, \ \  \mu_0   \,, \ \  \mu_1  \,, \ \ \mu_2  \,,  \ \  \Phi_1 \,, \ \  \Psi_1   \,, \ \  \Phi_2  \,, \ \ \Psi_2  \,, \ \  \Omega_0  \,, \ \  \Omega_1    \,, \ \  k \, \big\} \,.
  \label{functionlist}
\end{equation}
One can easily verify that this Ansatz is consistent with the equations of motion. 

We have arrived at exactly the same Ansatz as was used in  \cite{Ganchev:2021pgs,Ganchev:2021iwy,Ganchev:2021ewa}, but we imposed different symmetries to get to this point.  This is because the symmetries used in  \cite{Ganchev:2021pgs} are actually insufficient to remove two of the degrees of freedom that have been frozen in the Ansatz: (i) An independent function for $\chi_2(\xi)$ and (ii) The $d\xi$-component of $\tilde A^{12}$.   The Ansatz in \cite{Ganchev:2021pgs} could have allowed an arbitrary  complex function in $\chi_1+ i \chi_2$, and, while a gauge rotation can trivially remove  $\tilde A^{12}$, the result is then inconsistent with the  form of  $P$ and $\chi$ in (\ref{scalaransatz}).  
However the symmetries  used here are sufficient to this task. Nevertheless, the subsequent analysis  and results in    \cite{Ganchev:2021pgs,Ganchev:2021iwy,Ganchev:2021ewa} are correct because it was shown that the truncation is consistent with the equations of motion with $\chi_2 \equiv \tilde A^{12}_\xi   \equiv  0$.

We also note  that, in comparison to \cite{Ganchev:2021pgs,Ganchev:2021ewa,Ganchev:2021iwy}, we have absorbed the mode numbers and the frequencies  into the constant terms in $\Psi_1$ and $\Phi_1$. This is done via the  $SO(4)$ transformation:
\begin{equation}
U=\begin{pmatrix}
\cos\,(n\,\psi+t\,\omega) & -\sin\,(n\,\psi+t\,\omega) & 0 & 0\\
\sin\,(n\,\psi+t\,\omega) & \sin\,(n\,\psi+t\,\omega) & 0 & 0\\
0 & 0 & 1 & 0\\
0 & 0 & 0 & 1
\end{pmatrix} \,,
\end{equation}
and it acts on the gauge fields as:
\begin{equation}
\tilde{A}\rightarrow U\,\tilde{A}\,U^{-1}+\frac{1}{2\,g_0}\,(d\, U)\,U^{-1}.
\end{equation}
With the form of $U$ above, the gauge transformation induces shifts by $\frac12 \omega$ and $\frac12 n$ in respectively $\Phi_1$ and $\Psi_1$.  It also reduces the matrix, $P$, in \cite{Ganchev:2021pgs} to the diagonal form used here. As a result, the mode number and frequency are only implicit in our ansatz: they are constant terms in $\Phi_1$ and $\Psi_1$ that are set by the  boundary conditions of the gauge fields.

\subsection{The reduced action}
\label{sub:reduced-action}

The easiest way to express the equations of motion for this truncated system is to give the reduced action from which they can be derived.  This Lagrangian was given in \cite{Ganchev:2021pgs} and here we simply catalog this result. 

The bosonic Lagrangian can be decomposed into pieces:
\begin{equation}
    \cL ~=~ \cL_{\text{gravity}} + \cL_\chi + \cL_m + \cL_A + \cL_{CS} + \cL_Y - \sqrt{g}\, V \ ,
    \label{lagrangian_full}
\end{equation}
and the explicit expressions are: 
\begingroup
\allowdisplaybreaks
\begin{align}
\cL_{\text{gravity}} ~&\equiv~ -\frac14 \sqrt{g}\ R  \nonumber \\
~&=~  \frac{\Omega_1}{8\, g_0\, \xi}\, \bigg[ \,\frac{\Omega_1^2}{\Omega_0^2}\, \bigg(k' + \frac{2\,\xi}{1- \xi^2} \, k\bigg)^2  - 4\, \xi \,  \bigg(\partial_\xi\bigg( \xi\,  \frac{\Omega_0'}{\Omega_0}\bigg) + \frac{1}{\Omega_1} \,\partial_\xi\Big( \xi\,   \Omega_1' \Big)  \bigg)  -  \frac{16\, \xi^2}{(1- \xi^2)^2} \,    \bigg]
\\
\cL_\chi ~&\equiv~ \frac18 \sqrt{g}\, g^{\mu\nu}\, (\cD_\mu \chi_I)\, m^{IJ}\, (\cD_\nu \chi_J)  \nonumber \\
~&=~  \frac{e^{2(\mu_0 - \mu_1)}}{8\, g_0\, \xi \,(1- \xi^2) \,\Omega_1}\,  \,\bigg[\,  \Gamma \, \nu^2  ~-~   \xi^2\,(1- \xi^2)    \,\Omega_1^2 \, e^{-4\,\mu_0} \, \Big( \partial_\xi \Big (\sqrt{1- \xi^2}\, \nu \Big) \Big)^2    \, \bigg]
\\
\cL_m ~&\equiv~ \frac1{16} \sqrt{g} \, g^{\mu\nu} \, \Tr(m^{-1} (\cD_\mu m) m^{-1} (\cD_\nu m)) \nonumber   \\
~&=~  \frac{1}{2\, g_0\, \xi \,(1- \xi^2)^2 \,\Omega_1}\,  \,\bigg[\,  \Gamma \, \sinh^2 2\, \mu_0  ~-~   \xi^2\,(1- \xi^2)^2    \,\Omega_1^2 \,  \, \big( (\mu_0')^2 + (\mu_1')^2 +(\mu_2')^2 \big)   \, \bigg]
\\
\cL_A ~&\equiv~   -   \frac{1}{8}\, e \, g^{\mu \rho}  \, g^{\nu \sigma} \, m_{IK} \,m_{JL}\,  F_{\mu \nu }^{IJ}  \, F_{\rho \sigma }^{KL}  \nonumber  \\
~&=~  \frac1{2 g_0\, \xi\, \Omega_1} \Big[ \xi^2 \,\qty(e^{4 \mu_2}\ \Phi_1'^2 \,+\, e^{4\mu_1}\ \Phi_2'^2) \nonumber \\[-.3em]
& \qquad\qquad \qquad   -  \frac{\Omega_1^2}{\Omega_0^2}\,\qty(e^{4 \mu_2}\, \qty((1-\xi^2)\, \Psi_1' - k\, \Phi_1')^2 + e^{4\mu_1}\, \qty((1-\xi^2)\, \Psi_2' - k\, \Phi_2')^2) \Big]   
\\  
\cL_{CS} ~&\equiv~    \frac{1}{2}\,g_0 \, e  \, \varepsilon^{\mu \nu \rho} \,  \Big(A_\mu{}^{IJ}\, \partial_\nu  \widetilde A_\rho{}^{IJ}  ~+~\coeff{4}{3}\,  g_0 \, A_\mu{}^{IJ} \,  A_\nu{}^{JK}\, A_\rho{}^{KI} \,\Big)  \nonumber   \\
~&=~  \frac1{g_0} \qty( \Phi_1 \Psi_2' - \Psi_2 \Phi_1' + \Phi_2 \Psi_1' - \Psi_1 \Phi_2')  
\label{LCS}
\\
\cL_Y ~&\equiv~   \frac{1}{16}\,e  \, \varepsilon^{\mu \nu \rho} \,  {Y_\mu}{}^{IJ}  \, F_{\nu \rho}^{IJ}\nonumber \\
~&=~  \frac1{2 g_0} \qty(1-\xi^2)\, \nu^2\,\qty(\Psi_1\, \Phi_2' \,-\, \Phi_1 \Psi_2')   \\
V ~&=~ \frac{g_0^2}{2}\, e^{-4(\mu_1 + \mu_2)} \Big[1 - 2\, e^{2(\mu_1 + \mu_2)} \cosh(2\mu_0) + e^{4\mu_1} \sinh^2(2\mu_0) \nonumber \\
&\qquad\qquad \qquad \qquad +   \frac1{16}  \nu^2\, \qty(1-\xi^2) \qty((1-\xi^2)\,\nu^2 + 4\, e^{2(\mu_0 + \mu_1)}  - 8) \Big] \,,
\end{align}%
\endgroup
where, $'$ indicates a differentiation with respect to $\xi$.  We are also using the convenient shorthand that captures the mode dependence and minimal couplings: 
\begin{equation}
 \Gamma ~=~ 4\, \xi^2  \Omega_0^2 \, \Phi_1^2 ~-~ 2\,\Omega_1^2 \, \Big[\, (1-\xi^2 ) \,  \Psi_1 ~-~  k\, \Phi_1 \,\Big]^2 \,.
    \label{gauge_term}
\end{equation}
Again, in comparison to \cite{Ganchev:2021pgs,Ganchev:2021ewa,Ganchev:2021iwy}, we have absorbed superstratum/microstratum mode numbers and the frequencies  into the constant terms in $\Psi_1$ and $\Phi_1$.

There are also  three  integrals of the motion:
\begin{align}
\cI_0 ~\equiv~ & \frac{\xi^2 \, (1- \xi^2)}{k} \, \bigg[\,  \frac{1}{\xi\,\Omega_1} \, \big(  \,  \widehat \cL_{\text{gravity}} + \cL_A  + \sqrt{g}\, V -   \cL_\chi -   \cL_m\,\big)  \nonumber\\
& \qquad \qquad\quad -\frac{1}{g_0 }\,  \,\bigg( (\mu_0')^2 + (\mu_1')^2 +(\mu_2')^2  ~+~\frac{1}{4}\,  e^{-2(\mu_0 + \mu_1)}\, \Big( \partial_\xi \Big (\sqrt{1- \xi^2}\, \nu \Big) \Big)^2  \bigg)   \, \bigg]  \,,
\label{Ham-constr}
\end{align}
and 
\begin{align}
\cI_1 ~\equiv~  & \frac{e^{4\,  \mu_1}\,(1- \xi^2)\,\Omega_1}{\xi\,\Omega_0^2} \, \big( (1- \xi^2) \, \Psi_2' - k\, \Phi_2' \big)~-~  2\, \Phi_1  \, \big(1 - \coeff{1}{4}\, (1-\xi^2 ) \, \nu^2 \big) \,,  \label{Integral1}\\
\cI_2 ~\equiv~ &    \frac{e^{4\,  \mu_1}\, k\,\Omega_1}{\xi\,\Omega_0^2} \, \big( (1- \xi^2) \, \Psi_2' - k\, \Phi_2' \big) ~+~ \frac{e^{4\,  \mu_1}\,\xi \, \Phi_2' }{\Omega_1} ~-~ 2\, \Psi_1\, \big(1 - \coeff{1}{4}\, (1-\xi^2 ) \, \nu^2 \big) \,.\label{Integral2}
\end{align}
%

\subsection{The BPS equations}  
\label{ss:BPSequantions}

The BPS equations for this system were derived in \cite{Ganchev:2021iwy} and they start with algebraic constraints.  First one finds that $\Omega_1$ must be a constant.  Regularity requires that $\Omega_1$ is non-zero, and it can be re-scaled by scaling the time coordinate, $\tau$,  in (\ref{genmet1}).  For simplicity we take it to be\footnote{For AdS/CFT there is a more canonical normalization of time discussed in \cite{Ganchev:2021ewa}, namely $\lim\limits_{\xi\to1}\frac{g_{\tau\tau}}{g_{\sigma\sigma}}=-1$, where $\psi\to\tau+\sigma$. Rescaling time by $\tau \to s \,\tau$ implies the following for the fields: $\Omega_1 \to s \,\Omega_1$, $\Phi_i \to s \,\Phi_i$, and $k \to k/s$.}:
\begin{equation}
\Omega_1 ~\equiv~ 1 \,.
    \label{Omega1constraint}
\end{equation}
For non-trivial superstrata with $\nu \ne 0$ and $\mu_0 \ne 0$, the electric potentials are locked to the scalars:
\begin{equation}
    \Phi_1 ~=~ \frac12 e^{-2\mu_2}   \qand \Phi_2 ~=~ \frac12 (1 - e^{-2\mu_1}) \,,
    \label{Phipots}
\end{equation}
and the functions $\mu_0$  and $\nu$ are constrained by:
\begin{equation}
    (1-\xi^2) \,\nu^2 ~=~ \param1  \, \big(1-e^{4\mu_0}\big) \,,
    \label{numu0-2}
\end{equation}
for some constant, $\param1$.  This constant will prove to be a very important parameter in our new solutions because, in six dimensions, it determines the relative amounts of metric fluctuations (represented by $\mu_0$) and tensor multiplet-fluctuations (represented by $\nu$).

The remaining independent    first-order BPS equations determine $\mu_0$, $\Psi_1$, $\Psi_2$, $\Omega_0$ and $k$:
\begin{align}
  \xi \, \partial_\xi \mu_0  ~-~  2\,\sinh(2 \mu_0) \, \qty[\Psi_1~-~ \frac{k}{ (1-\xi^2) } \, \Phi_1]  &   ~=~0 
    \label{eq:res_mu_0-2} \,, \\   
 \qty( \partial_\xi\Psi_1-\frac{ k}{(1-\xi^2)} \partial_\xi\Phi_1 )   ~+~ \frac{\xi\, \Omega_0^2}{ (1-\xi^2)^2}  \, e^{-2\mu_1 - 4\mu_2}\,  \bigg(1 - e^{2\mu_1} \cosh(2 \mu_0) - \frac{1-\xi^2}4 \nu^2  \bigg) &   ~=~0 
    \label{eq:res_nu_mu0-2} \,,  \\
 \qty(  \partial_\xi\Psi_2 - \frac{ k}{(1-\xi^2)} \partial_\xi\Phi_2 )   ~-~ \frac{\xi\, \Omega_0^2}{ (1-\xi^2)^2}  \, e^{-4\mu_1 - 2 \mu_2}\,  \bigg( 1 - e^{2\mu_2} - \frac{1-\xi^2}4 \nu^2 \bigg) &   ~=~0 \,, \label{eq:res_gauge4-2} \\
     \xi \partial_\xi \big(\ln\Omega_0 - (\mu_1 + \mu_2)\big)   ~+~  2\,\cosh(2\mu_0) \, \qty[\Psi_1~-~ \frac{k}{ (1-\xi^2) } \, \Phi_1] +  \frac{2\,\xi^2}{1-\xi^2} &  \nonumber \\
    \qquad  -  2\, \bigg( \Psi_2 -\frac{k}{(1-\xi^2)} \qty(\Phi _2  - \frac12   ) \bigg)    &   ~=~0  \,,
    \label{eq:res_omega_0-2}  \\
\partial_\xi \bigg( \frac{k}{(1-\xi^2)}\bigg) ~+~\frac{2\, \xi\, \Omega_0^2}{ (1-\xi^2)^2} \,e^{-2(\mu_1 + \mu_2)} \,  \bigg(  1   -  e^{2\mu_2}  - e^{2\mu_1} \cosh(2 \mu_0) - \frac{1-\xi^2}4 \nu^2 \bigg)&   ~=~0 \,.
    \label{eq:res_k-2} 
\end{align}

The last two scalar functions,  $\mu_1$ and $\mu_2$, are not determined by the BPS equations. To this end, we introduce the quantity:
\begin{equation}
 \cI ~\equiv~  e^{2 \mu_1} \xi\, \partial_\xi \mu_1 ~-~  \frac{k}{(1-\xi^2)} ~-~2 \, \bigg( \Psi_1 ~-~ \frac{ k}{ (1-\xi^2) } \, \Phi_1\bigg)\,  \bigg(  1   - \frac{1}{4}\,(1-\xi^2) \nu^2 \bigg) \,,
    \label{integral1}
\end{equation}
which one can show is conserved,
\begin{equation}
\partial_\xi \cI ~=~  0\,,
\label{eom1}
\end{equation}
under the equations of motion from  \cite{Ganchev:2021pgs}. The latter also give us 
\begin{align}
\frac{1}{2}\,\partial_\xi & \Big[\, \xi  \partial_\xi \,\big(  e^{2 \mu_2} \big) \Big]  \nonumber \\ ~-~ &\frac{2\,\xi\, \Omega_0^2\,e^{-2  \mu_2 }}{ (1-\xi^2)^2} \,  \bigg[ \,1 -  e^{-2 (\mu_0+\mu_1)} \, \bigg(   e^{4 \mu_0}~-~ \frac{1}{2} e^{2 \mu_2} \,(1+ e^{4 \mu_0})~+~ \bigg(1- \frac{1}{4}\,(1-\xi^2) \nu^2 \bigg) \bigg)  \bigg]~=~  0\,.
    \label{eom2}
\end{align}
%

\subsection{The superstratum solution} 
\label{ss:superstratum}

It is  useful to note that the ``single-mode'' $(1,0,n)$ superstratum \cite{Mayerson:2020tcl, Ganchev:2021pgs} corresponds to: 
\begin{equation}
\begin{aligned}
\nu   ~=~&\alpha_0 \,   \xi^n  \,, \qquad  \mu_1 ~=~ \coeff{1}{2}\,  \log \Big[ \, 1 - \coeff{1}{4}\, \alpha_0^2 \,  (1-\xi^2)\, \xi^{2n}  \, \Big]   \,,\qquad     \mu_0 ~=~ \mu_2 ~=~ 0 \,,  \\
  \Phi_1  ~=~  &\frac{1}{2}  \,, \qquad \Psi_1  ~=~ \frac{n}{2}    \,,  \\  
  \Phi_2 ~=~ &\frac{1}{2} \,\bigg[\,1 ~-~  \frac{1}{\big(\,1 - \coeff{1}{4}\, \alpha_0^2 \, (1-\xi^2)\,  \xi^{2n} \, \big)}  \, \bigg]\,, \qquad  \Psi_2 ~=~
-   \frac{\alpha_0^2 }{8}  \,\frac{ \xi^{2n+2}}{\big(\, 1 - \coeff{1}{4}\, \alpha_0^2 \,  (1-\xi^2)\, \xi^{2n} \, \big)} \,,\\ 
  \Omega_0  ~=~ &  \sqrt{\, 1 - \coeff{1}{4}\, \alpha_0^2 \,  (1-\xi^2)\, \xi^{2n} }     \,, \qquad  \Omega_1 ~=~  1    \,, \qquad   k ~=~\xi^2  \,.
\end{aligned}
  \label{ssres1}
\end{equation}
where we have made a gauge choices for $ \tilde A^{34}$, (\ref{gauge_ansatz}), so that $\Phi_2(1) =0$ and $\Psi_2(0) = 0$. Note also that the mode number, $n\in\mathbb{N}$, of the superstratum appears in the constant value of $\Psi_1$, and in the notation of \cite{Ganchev:2021pgs,Ganchev:2021ewa,Ganchev:2021iwy}, we have $\omega=0$, hence the value of $\Phi_1$.

\subsection{A  solution on the ``special locus''} 
\label{sub:a_special_locus}

The ``special locus'' first emerged in the study of microstrata  \cite{Ganchev:2021ewa,Ganchev:2021iwy}, where it became necessary to restrict to it in order to have regular, BPS and non-BPS solutions.  This locus is defined by $\mu_0 \equiv \mu_1$, $\mu_2 \equiv 0$ and  $\param1 =2$.   This subsequently led to a new family of BPS solutions   \cite{Ganchev:2021iwy} with:
\begin{equation}
\Omega_1 ~\equiv~ 1 \,, \quad \param1  ~=~ 2 \,,  \quad   \mu_0 ~\equiv~ \mu_1 \,,\quad \mu_2 ~\equiv~ 0\,, \quad  \Phi_1 ~\equiv~\frac{1}{2}\,,  \quad \Psi_1 ~\equiv~ \frac p2\,, \quad\Phi_2 ~=~ \frac12 (1 - e^{-2\mu_0})  \,,
    \label{specialmu}
\end{equation}
for all $\xi$.

The equations for $\mu_0$, $\Psi_2$, $\Omega_0$ and $k$ then reduce to:
\begin{align}
 \frac{ \xi \, \partial_\xi \mu_0}{ \sinh(2 \mu_0) }   &   ~=~ p ~-~ \frac{k}{ (1-\xi^2) } 
    \label{eq:res_mu_0-3} \,,  \\
\partial_\xi\Psi_2 + \frac{ k}{2\,(1-\xi^2)} \partial_\xi(e^{-2\mu_0})   &  ~=~  \frac{\xi\, \Omega_0^2}{ (1-\xi^2)^2}  \,  e^{-2\mu_0 } \sinh(2\mu_0)   \,, \label{eq:res_gauge4-3} \\
     \xi \partial_\xi \big(\ln\Omega_0 -  \mu_0  \big)  & ~=~ - \cosh(2\mu_0) \, \bigg[\, p ~-~ \frac{k}{ (1-\xi^2) } \bigg] - \frac{2\,\xi^2}{1-\xi^2}   \nonumber \\
& \qquad\qquad    ~+~ 2 \Psi_2~+~  \frac{ k}{(1-\xi^2)}\, e^{-2\mu_0}\,,
    \label{eq:res_omega_0-3}  \\
\partial_\xi \bigg( \frac{k}{(1-\xi^2)}\bigg) & ~=~  \frac{2\, \xi\,e^{-2\mu_0 } \,\Omega_0^2}{ (1-\xi^2)^2}   \,.
    \label{eq:res_k-3} 
\end{align}

It was also shown in  \cite{Ganchev:2021iwy}  that this system of equations admits an analytic family of solutions, parametrized by a real constant $\gamma$. Define
\begin{equation}
    \Lambda_1^2 ~\equiv~ 1 - \gamma^4 \,\xi^{4p+2} \ ,\quad
    \Lambda_2^2 ~\equiv~  (2p+1)\, \gamma^2\, \xi^{2p}\, (1-\xi^2)
\end{equation}
and  the solutions are then given by 
\begin{equation}
\begin{aligned}
    &\nu ~=~ \frac{2\sqrt{2}}{\sqrt{1-\xi^2}} \frac{\Lambda_1\Lambda_2}{\Lambda_1^2 + \Lambda_2^2} \ ,
    \qquad
    \mu_0 ~=~ \mu_1 ~=~ -\arctanh\qty(\frac{\Lambda_2^2}{\Lambda_1^2}) \ ,
    \qquad 
    \mu_2 ~=~ 0 \ ,
    \\
    &\Phi_1 ~=~ \frac12 \ ,
    \qquad\qquad
    \Psi_1 ~=~ \frac{p}{2}  \ ,
    \\
    &\Phi_2 ~=~ - \frac{\Lambda_2^2}{\Lambda_1^2 - \Lambda_2^2} \ ,
    \qquad
    \Psi_2 ~=~ - \frac{\xi^2}{1-\xi^2} \frac{\Lambda_2^2}{\Lambda_1^2 - \Lambda_2^2} \qty(1 - \gamma^2 \xi^{2p}) \ ,
    \\
    &k ~=~ \xi^2 \qty(1 \,-\, \gamma^2\, \xi^{2p}\, \frac{\Lambda_2^2}{\Lambda_1^2}) \ ,
    \qquad
    \Omega_0 ~=~ 1 - \frac{\Lambda_2^2}{\Lambda_1^2} \ ,
    \qquad\qquad
    \Omega_1 ~=~ 1 \ .
\end{aligned}
\label{eq:analytic_locus}
\end{equation}
for general $p\in\mathbb{N}$.

If one then defines 
\begin{equation}
    \alpha ~\equiv~ \sqrt{8(2n+1)} \,\gamma\,,\quad p\rightarrow n\,,
    \label{eq:alpha_gamma_conv}
\end{equation}
one has $\nu \sim \alpha \, \xi^n$ as $\xi \to 0$, and these solutions match with the perturbative expansion and the numerical solutions found in \cite{Ganchev:2021pgs} (for $n=2$, $\omega_0=0$).

Using the form of the metric (\ref{genmet1}), one can compute explicitly the coefficient of $d\psi^2$ in the metric, and find the range of parameters for which the solution is CTC-free. We  find that this requires
\begin{equation}
\gamma^2 \leq \frac1{4 p + 1} \,.
\label{eq:ctc_limit_special_locus}
\end{equation}
When $\gamma$ saturates this bound, the solution is asymptotic to AdS$_2 \times$ S$^1$.

\section{Solving the BPS equations in general} 
\label{sec:genBPS}

\subsection{The BPS layers} 
\label{ss:BPSlayers}

We start by using the algebraic constraints (\ref{Omega1constraint}),  (\ref{Phipots})   and    (\ref{numu0-2}) in (\ref{eq:res_mu_0-2})--(\ref{eq:res_k-2}) and introducing the functions:
\begin{equation}
\begin{aligned}
F(\xi) ~\equiv~&   2\, \Psi_1 ~-~ \frac{  2\, k}{ (1-\xi^2) } \, \Phi_1 \,,  \\
G(\xi) ~\equiv~& 2\,  \Psi_2 ~+~ 1 ~-~ \frac{2\,  k}{(1-\xi^2)} \Big(\Phi _2  - \frac12 \Big) \,,  \\
H(\xi) ~\equiv~&  \frac{2\, \xi^2\, \Omega_0^2\,e^{-2(\mu_1 + \mu_2)}}{ (1-\xi^2)^2}  \,.
\end{aligned}
    \label{FGHdefns}
\end{equation}
By taking linear combinations of  (\ref{eq:res_k-2}) with  (\ref{eq:res_mu_0-2})--(\ref{eq:res_omega_0-2}), one finds that the latter four equations can be written:
\begin{equation}
\begin{aligned}
 & \xi \,\partial_\xi \, F  ~=~   - H \,, \qquad   \xi \,\partial_\xi \, G  ~=~  H \, \cosh(2 \mu_0) \,,  \qquad 
\frac{\xi \,\partial_\xi \mu_0}{\sinh(2 \mu_0)}~=~ F \,, \\
& \xi \,\partial_\xi \, \log \big[ H \, \sinh(2 \mu_0)\big] ~=~    2\, G  \,.
\end{aligned}
    \label{BaseSystem}
\end{equation}

Observe that this is a  ``base-layer'' - a system  of four first-order, non-linear equations for the four functions: $F,G,H$ and $\mu_0$. This system has been decoupled from the other BPS equations and from the equations of motion.  As we will see in Section \ref{sec:uplift}, this is the first-order, non-linear system that determines the base geometry in six dimensions,

Now observe that if one uses (\ref{eq:res_k-2}) to eliminate the derivatives of $k$ in  (\ref{eom1}) and then takes the sums and differences with  (\ref{eom2}), one finds:
\begin{equation}
\begin{aligned}
\frac{1}{2}\,   \xi \,\partial_\xi & \, \Big[ \xi \,\partial_\xi \, \big (e^{2\mu_1}  + e^{2\mu_2}   \big)  \Big] ~-~ 2\, H  \, \cosh^2 \mu_0 \,  \big( e^{2\mu_1}  +  e^{2\mu_2}   \big) \\ 
 & =~ -  H \, e^{2\mu_0} ~-~   H \, \big ( 1+ e^{-2\mu_0}  \big) \, \Big ( 1- \coeff{1}{4}\,(1-\xi^2)\, \nu^2 \Big)    ~+~   \xi \,\partial_\xi \, \Big[ F\,\, \Big ( 1- \coeff{1}{4}\,(1-\xi^2)\, \nu^2 \Big)  \Big]   \,, \\
\frac{1}{2}\,  \xi \,\partial_\xi &\, \Big[ \xi \,\partial_\xi \, \big (e^{2\mu_1}  -  e^{2\mu_2}   \big)  \Big] ~-~ 2\, H  \, \sinh^2 \mu_0 \,  \big( e^{2\mu_1}  -  e^{2\mu_2}   \big) \\ 
 &= ~    H \, e^{2\mu_0} ~-~   H \, \big ( 1- e^{-2\mu_0}  \big) \,  \Big ( 1- \coeff{1}{4}\,(1-\xi^2)\, \nu^2 \Big) ~+~ \xi \,\partial_\xi \, \Big[ F\,  \Big ( 1- \coeff{1}{4}\,(1-\xi^2)\, \nu^2 \Big)  \Big]   \,, 
\end{aligned}
    \label{Layer1}
\end{equation}

Since  $F,G,H$ and $\mu_0$ are known from solving the base-layer equations,  (\ref{BaseSystem}),  the functions appearing in $\eqref{Layer1}$ are known, with $\nu$ being determined by (\ref{numu0-2}).  Therefore the equations in this ``first layer,'' (\ref{Layer1}), are {\it linear} equations in $(e^{2\mu_1}  +  e^{2\mu_2})$ and $(e^{2\mu_1}  -  e^{2\mu_2})$.  The solutions to the base layer determine  the sources and coefficients of this linear system, and the functions determined by these equations are precisely the electrostatic potentials  (\ref{Phipots}).  

Once this linear system is solved, one finally determines $k$ from the first-order equation (\ref{eq:res_k-2}), which may be written as a linear ``second-layer'' equation:
\begin{equation}
\xi \partial_\xi\, \bigg( \frac{k}{(1-\xi^2)}\bigg) ~=~ H\, \bigg[  e^{2\mu_2}  +  e^{2\mu_1}\, \cosh(2 \mu_0) -  \Big ( 1- \coeff{1}{4}\,(1-\xi^2)\, \nu^2 \Big) \bigg] \,,
\label{Layer2}
\end{equation}
where the sources are entirely determined by the previous layers.

The BPS equations obtained in \cite{Ganchev:2021iwy} have precisely the ``linear structure'' discovered in the earlier BPS systems that underpin microstate geometries \cite{Bena:2004de,Bena:2011dd,Bena:2015bea}.  Note also that the base geometry, which is determined by  $F,G,H$ and $\mu_0$ , is entirely independent of $\nu$, or $\param1$, and hence independent of the amplitude of the standard superstratum excitation.

\subsection{Solving the base layer}  
\label{ss:baselayersol}

The base-layer system, (\ref{BaseSystem}), has a remarkable structure that enables us to solve the system completely.  Indeed, one can easily verify that these equations imply:
\begin{equation}
  \xi \,\partial_\xi \, \Big[ \xi \,\partial_\xi \,  (F+G)   ~-~  (F+G)^2  \Big]   ~=~  0   \,,   \qquad      \xi \,\partial_\xi \, \Big[ \xi \,\partial_\xi \,  (F-G)   ~+~  (F-G)^2  \Big]   ~=~  0   \,.
  \label{AmazingEqs}
\end{equation}
These equations are trivially solved to give:
\begin{equation}
F   ~=~  \frac{1}{2} \,\bigg[q_1 \frac{1+ \gamma_1 \, \xi^{2q_1}}{1- \gamma_1 \, \xi^{2q_1}} ~-~  q_2 \frac{1+ \gamma_2 \, \xi^{2q_2}}{1- \gamma_2 \, \xi^{2q_2}} \bigg]    \,, \qquad G   ~=~  \frac{1}{2} \,\bigg[ q_1 \frac{1+ \gamma_1 \, \xi^{2q_1}}{1- \gamma_1 \, \xi^{2q_1}} ~+~  q_2 \frac{1+ \gamma_2 \, \xi^{2q_2}}{1- \gamma_2 \, \xi^{2q_2}} \bigg]   \,,
  \label{FGres1}
\end{equation}
where the four constants of integration in   (\ref{AmazingEqs}) are $q_1, q_2, \gamma_1$ and $\gamma_2$

One should note that with these choices one has
\begin{equation}
 \xi \,\partial_\xi \,  (F+G)   ~-~  (F+G)^2    ~=~  - q_1^2   \,,   \qquad       \xi \,\partial_\xi \,  (F-G)   ~+~  (F-G)^2    ~=~  +q_2^2   \,.
  \label{IntConst1}
\end{equation}
Note that we have implicitly chosen the signs of the constants of integration in these equations.  One can choose the opposite sign, and this leads to imaginary exponents and solutions with singular behavior at $\xi=0$.  Our choice leads to rational functions, and one must have $2q_1,\,2q_2 \in \ZZ$ if one wishes to avoid branch cuts.

From  the third equation in  (\ref{FGres1}) and  (\ref{BaseSystem}) one obtains:
\begin{equation}
\mu_0 ~=~ - \frac{1}{2}\, \log \bigg[   \frac{\xi^{q_2} \, (1- \gamma_1 \, \xi^{2 q_1}) ~-~ \beta\, \xi^{q_1} \,  (1- \gamma_2 \, \xi^{2 q_2})  }{\xi^{q_2} \, (1- \gamma_1 \, \xi^{2 q_1}) ~+~ \beta\, \xi^{q_1} \,  (1- \gamma_2 \, \xi^{2q_2}) } \bigg]       \,.
  \label{mu0res1}
\end{equation}
where $\beta$ is yet another constant of integration.

Substituting these back into  (\ref{BaseSystem}) one finds that the first order system yields:
\begin{equation}
H  ~=~  \frac{2 \, q_2^2 \, \gamma_2 }{(1- \gamma_1 \, \xi^{2q_1})^2\, (1- \gamma_2 \, \xi^{2q_2})^2  } \Big(  \xi^{2q_2} \, (1- \gamma_1 \, \xi^{2q_1})^2 ~-~ \beta^2 \, \xi^{2 q_1} \,  (1- \gamma_2 \, \xi^{2q_2})^2   \Big)    \,.
  \label{Hres1}
\end{equation}
along with a constraint on the constants of integration:
\begin{equation}
 q_1^2 \gamma_1 ~-~   \beta^2\,  q_2^2 \, \gamma_2  ~=~ 0\,,
  \label{constraint1}
\end{equation}
which, in particular,  means that $\gamma_1$ and  $\gamma_2$ must have the same sign.   

In Section \ref{ss:Regularity}, we will constrain some of these parameters through re-parametrizations and regularity. For the present we will construct the rest of the solution in general form.

\subsection{Solving the remaining layers of BPS equations}  
\label{ss:layer1sol}

One can actually solve the linear system of the first layer,  (\ref{Layer1}), by quadrature.  Indeed, it follows from   (\ref{AmazingEqs}) that for any function, $P$, one has:
\begin{equation}
\xi \,\partial_\xi \, \Big[ (F \pm G)^2 \,  \xi \,\partial_\xi \, \big ( (F \pm G)^{-1} \,P  \big)  \Big]  ~=~  (F \pm G) \,\Big[   \xi \,\partial_\xi \, \big (\xi \,\partial_\xi \,  P  \big) ~\mp~ 2\, \big(  \xi \,\partial_\xi  (F \pm G) \big) \, P \Big] \,.
\label{simpeqn}
\end{equation}
Now observe that 
\begin{equation}
 2\, \big(  \xi \,\partial_\xi  (F +  G) \big)   ~=~4\, H\,  \sinh^2(\mu_0) \,, \qquad  2\, \big(  \xi \,\partial_\xi  (F -  G) \big)   ~=~ 4\, H\,  \cosh^2(\mu_0)   \,,
\end{equation}
which means that the differential operators on the left-hand sides of (\ref{Layer1}) can be rewritten using (\ref{simpeqn}), to give the following form of the first layer:
\begin{equation}
\begin{aligned}
\xi \,\partial_\xi \, \Big[ & (F - G)^2  \,  \xi \,\partial_\xi \, \big ( (F - G)^{-1} \,\big (e^{2\mu_1}  +  e^{2\mu_2}   \big)   \big)  \Big]   \\ 
 & ~=~  (F - G)\,\bigg[ -2 \, H \, e^{2\mu_0} ~+~  2\, \xi \,\partial_\xi \, \Big[ F\,\, \Big ( 1- \coeff{1}{4}\,(1-\xi^2)\, \nu^2 \Big)  \Big]  \bigg]  \,, \\
 \xi \,\partial_\xi \, \Big[& (F + G)^2 \, \xi \,\partial_\xi \, \big ( (F + G)^{-1} \,\big (e^{2\mu_1}  -  e^{2\mu_2}   \big)   \big)  \Big]  \\ 
 & ~=~ (F + G)\,\bigg[   2 \, H \, e^{2\mu_0} ~-~ 4 \, H \, \Big ( 1- \coeff{1}{4}\,(1-\xi^2)\, \nu^2 \Big) ~+~  2\, \xi \,\partial_\xi \, \Big[ F\,  \Big ( 1- \coeff{1}{4}\,(1-\xi^2)\, \nu^2 \Big)  \Big]  \bigg]  \,, 
\end{aligned}
    \label{Layer1b}
\end{equation}

The important point is that the sources on the right-hand side of these expressions consist of known functions given by    (\ref{numu0-2}), (\ref{FGres1}), (\ref{mu0res1}) and  (\ref{Hres1}).  One can therefore solve these equations by integrating and we find:
\begin{equation}
\begin{aligned}
\big (e^{2\mu_1}  +  e^{2\mu_2}   \big)     ~=~  & \param1 \, \frac{ \beta\, \xi^{q_1} \,  (1- \gamma_2 \, \xi^{2q_2}) }{\xi^{q_2} \, (1- \gamma_1 \, \xi^{2q_1}) ~-~ \beta\, \xi^{q_1} \,  (1- \gamma_2 \, \xi^{2q_2})}  ~-~ \frac{2\,c_4 }{ 1- \gamma_2 \, \xi^{2q_2}}    \\ 
& ~-~\frac{1+ \gamma_2 \, \xi^{2q_2} }{ 1- \gamma_2 \, \xi^{2q_2}}  \big(  (c_4+2)\,q_2\, \log \xi    ~+~ c_6 \big) \,,
\\ 
\big (e^{2\mu_1}  -  e^{2\mu_2}   \big)     ~=~  & \param1 \, \frac{ \xi^{q_2} \, (1- \gamma_1 \, \xi^{2q_1}) }{\xi^{q_2} \, (1- \gamma_1 \, \xi^{2q_1}) ~-~ \beta\, \xi^{q_1} \,  (1- \gamma_2 \, \xi^{2q_2})} ~-~  \frac{2\,c_5 }{ 1- \gamma_1 \, \xi^{2q_1}}   \\ 
&  ~-~ \frac{1+ \gamma_1 \, \xi^{2q_1} }{ 1- \gamma_1 \, \xi^{2q_1}}  \big( (c_5 -2)\, q_1\,\log \xi + c_7\big )   \,,
\end{aligned}
    \label{Layer1bsol1}
\end{equation}
 where $c_4,c_5,c_6$ and $c_7$ are constants of integration.
 
Exponentiating  (\ref{mu0res1}) yields:
\begin{equation}
e^{2\, \mu_0 }~=~ \frac{\xi^{q_2} \, (1- \gamma_1 \, \xi^{2q_1}) ~+~ \beta\, \xi^{q_1} \,  (1- \gamma_2 \, \xi^{2q_2}) }{\xi^{q_2} \, (1- \gamma_1 \, \xi^{2q_1}) ~-~ \beta\, \xi^{q_1} \,  (1- \gamma_2 \, \xi^{2q_2})  } \,,
  \label{mu0res12}
\end{equation}
and hence 
\begin{equation}
\begin{aligned}
\big (\coeff{1}{2}\, \param1\,  e^{2\mu_0}  - e^{2\mu_1}   \big)     ~=~&     \frac{1}{2} \, \frac{1+ \gamma_2 \, \xi^{2q_2} }{ 1- \gamma_2 \, \xi^{2q_2}}  \big(  (c_4+2)\,q_2\, \log \xi    + c_6 \big)  ~+~  \frac{c_4 }{ 1- \gamma_2 \, \xi^{2q_2}}  \\ 
& ~+~ \frac{1}{2} \, \frac{1+ \gamma_1 \, \xi^{2q_1} }{ 1- \gamma_1 \, \xi^{2q_1}}  \big( (c_5 -2)\, q_1\,\log \xi + c_7\big ) ~+~ \frac{c_5 }{ 1- \gamma_1 \, \xi^{2q_1}}  \,,
\\ 
 e^{2\mu_2}    ~=~ &  - \coeff{1}{2}\,  \param1  ~-~  \frac{1}{2} \, \frac{1+ \gamma_2 \, \xi^{2q_2} }{ 1- \gamma_2 \, \xi^{2q_2}}  \big(  (c_4+2)\,q_2\, \log \xi  + c_6 \big)  ~-~  \frac{c_4 }{ 1- \gamma_2 \, \xi^{2q_2}}  \\ 
& ~+~ \frac{1}{2} \, \frac{1+ \gamma_1 \, \xi^{2q_1} }{ 1- \gamma_1 \, \xi^{2q_1}}  \big( (c_5 -2)\, q_1\,\log \xi + c_7\big ) ~+~ \frac{c_5 }{ 1- \gamma_1 \, \xi^{2q_1}}     \,.
\end{aligned}
    \label{Layer1bsol2}
\end{equation}

One can now integrate the last layer, (\ref{Layer2}), to arrive at:
\begin{equation}
\begin{aligned}
k   ~=~     -(1-\xi^2)\, \bigg[ \,&    \big(  (c_4+2)\,q_2\, \log \xi +    c_4 +c_6 \big)  \, \frac{q_2\, \gamma_2 \, \xi^{2q_2}}{(1- \gamma_2 \, \xi^{2q_2})^2}~+~ \frac{1}{2}\,  \frac{c_4\, q_2}{1- \gamma_2 \, \xi^{2q_2}} \\
&+~  \big(  (c_5-2)\,q_1\, \log \xi +    c_5 +c_7 \big)  \, \frac{q_1\, \gamma_1 \, \xi^{2q_1}}{(1- \gamma_1 \, \xi^{2q_1})^2}~+~ \frac{1}{2}\,  \frac{c_5\, q_1}{1- \gamma_1 \, \xi^{2q_1}}~+~c_8\,  \bigg]   \,,
\end{aligned}
    \label{ksol1}
\end{equation}
where $c_8$ is a constant of integration.

The rest of the fields can now be easily determined algebraically using \eqref{Omega1constraint},  \eqref{Phipots}, \eqref{numu0-2} and \eqref{BaseSystem}. For completeness, the expressions are given in Appendix \ref{app:restSols}.

\subsection{Gauge fixing, regularity and asymptotic behaviour}
\label{ss:Regularity}

We start by removing redundancies from the parametrization of the solution and ensuring the regularity of the base-layer, (\ref{BaseSystem}), functions $F,\,G,\,H$ and $\mu_0$ for  $0 \le \xi <1$. 

\subsubsection{The base system}
\label{ss:basesystem}

Note that:
\begin{itemize}
 \item The  solution to the base-layer, first-order BPS system is invariant under both $(q_1 \to- q_1 , \gamma_1\to1/\gamma_1, \beta\to -\beta/\gamma_1)$ and $(q_2 \to- q_2, \gamma_2\to1/\gamma_2, \beta\to -\beta/\gamma_2)$, which means that we can choose $q_1 \geq 0$ and $q_2 \geq 0$.
 \item The regularity of $\mu_0$ as $\xi \to 0$ implies  $q_1 \geq q_2$.
 \item The absence of any singularity in $F$ and $G$ (except possibly at $\xi=1$, due to the finiteness of $k$ at infinity and the form of \eqref{FGHdefns}) implies $\gamma_1, \gamma_2 \leq 1$.
\item If $\param1 \ne 0$, then regularity of $\nu$  at infinity ($\xi =1$)  means that $\mu_0 \to 0$ as $\xi \to 1$.  This means either $\gamma_1 = 1$ or $\gamma_2 = 1$ or $\beta = 0$. The choice $\beta = 0$ will be examined later. When choosing between the gammas, positivity of $e^{-2\mu_0}$ implies one must take $\gamma_2 = 1$.   We therefore have either $\param1 = 0$ or $\gamma_2 = 1$, for $\beta\neq0$.
\end{itemize}

Next, we observe that there is an unfixed coordinate choice: $\xi \to \xi^\lambda$, with $\lambda >0$.  This leaves $\xi =0,1$ unchanged.  Moreover, the BPS equations    (\ref{BaseSystem}) ,  (\ref{Layer1}) and (\ref{Layer2}) only involve the differential operator $\xi \partial_\xi$, and  this scales by $\lambda^{-1}$.  The system is therefore invariant under this re-definition, as long as, one leaves the $\mu_j$ unchanged but replaces: 
\begin{equation}
F\,,  \, G ~\rightarrow~ \lambda^{-1} F\,, \,  \lambda^{-1} G  \,, \quad H~\rightarrow~ \lambda^{-2} H  \,,  \quad \kappa(\xi)  ~\rightarrow~   \lambda^{-1}  \kappa(\xi) \,,
  \label{scales}
\end{equation}
where $\kappa(\xi)$ is defined in (\ref{kappadefn}). We therefore fix the rescaling of $\xi$ by setting $q_2 = 1$ in   (\ref{IntConst1}).
  
To summarize: for $\param1 \ne 0 $ , and using   (\ref{constraint1}), we arrive at:
\begin{equation}
\param1 \ne 0 \,, \qquad q_2 ~=~ 1\,, \qquad 0 ~\le~ \gamma_1  ~\le~ 1 \,, \qquad \gamma_2  ~=~ 1 \,,   \qquad  \beta^2 ~=~ q_1^2 \gamma_1\,;
  \label{reduction1}
\end{equation}
while for $\param1 =0$, one simply has  $\gamma_1, \gamma_2 \leq 1$ and
\begin{equation}
\param1 ~=~ 0 \,, \qquad   q_2 ~=~ 1\,,   \qquad  q_1^2 \gamma_1 ~=~   \beta^2\, \gamma_2 \,. 
  \label{reduction3}
\end{equation}
We will discuss this solution in more detail in Section \ref{sec:Example}. Its significance lies in the fact that $\nu\equiv 0$, and yet it can produce a scaling superstratum.  This can then be S-dualized to a scaling, pure NS superstratum.

\subsubsection{Smoothness of the complete solution}
\label{ss:smoothness}

To obtain a smooth solution one must remove the $\log$ terms  in the solutions for $\mu_1$, $\mu_2$ and $k$ and require  $k \to 0$ as $\xi \to 0$. We will do this without implementing the constraints from the previous section and comment on how they combine at the end. This means one must take: 
\begin{equation}
c_4 ~=~ -2\,, \qquad c_5 ~=~ 2\,,  \qquad c_8 ~=~ q_2 - q_1 \,,
    \label{smooth1}
\end{equation}
With these choices one has:
\begin{equation}
\begin{aligned}
& \coeff{1}{2}\, \param1\,  e^{2\mu_0}  - e^{2\mu_1}     ~=~   - \coeff{1}{2}\,  (c_6 + c_7)  ~+~  \frac{c_6- 2 }{ 1- \gamma_2 \, \xi^{2q_2}}
~+~ \frac{c_7+2  }{ 1- \gamma_1 \, \xi^{2q_1}}  \,,
\\ 
 & e^{2\mu_2}    ~=~  - \coeff{1}{2}\,  (\param1-c_6 + c_7)    ~-~  \frac{c_6- 2}{ 1- \gamma_2 \, \xi^{2q_2}}  ~+~ \frac{c_7+2}{ 1- \gamma_1 \, \xi^{2q_1}}     \,.
\end{aligned}
    \label{Layer1bsol3}
\end{equation}
and
\begin{equation}
\frac{k}{1-\xi^2}   ~=~     - ( c_6 -2)  \, \frac{q_2\, \gamma_2 \, \xi^{2q_2}}{(1- \gamma_2 \, \xi^{2q_2})^2}~+~ \frac{q_2\,\gamma_2 \, \xi^{2q_2}}{1- \gamma_2 \, \xi^{2q_2}}  
~-~  (c_7+2 )  \, \frac{q_1\, \gamma_1 \, \xi^{2q_1}}{(1- \gamma_1 \, \xi^{2q_1})^2}~-~ \frac{ q_1\, \gamma_1 \, \xi^{2q_1}}{1- \gamma_1 \, \xi^{2q_1}}  \,.
    \label{ksol2}
\end{equation}

The parameters $\gamma_2$, $c_6$  and $c_7$  determine the values of the scalars at  infinity ($\xi =1$). If one  requires that the solution goes to the supersymmetric critical point at infinity, where all the $\mu_j$ vanish, one must take: 
\begin{equation}
\gamma_2 ~=~ 1\,, \qquad c_6 ~=~ 2\,,  \qquad c_7 ~=~  \frac{(1-\gamma_1)\, \param1 - 4}{1+\gamma_1}\,.
    \label{susyinf}
\end{equation}
This leads to:
\begin{equation}
e^{2\mu_0} ~=~   \frac{\xi^{q_2} \, (1- \gamma_1 \, \xi^{2q_1}) ~+~ \beta\, \xi^{q_1} \,  (1-  \xi^{2q_2})  }{\xi^{q_2} \, (1- \gamma_1 \, \xi^{2q_1}) ~-~ \beta\, \xi^{q_1} \,  (1-  \xi^{2q_2}) }      \,.
  \label{mu0res2}
\end{equation}
\begin{equation}
\begin{aligned}
\big( e^{2\mu_1} ~-~ \coeff{1}{2}\, \param1 \,e^{2\mu_0}  \big)     ~=~&   - \frac{1}{2} \,(\param1-2) \, \frac{(1 -\gamma_1) }{ (1 + \gamma_1)}\, \frac{(1+ \gamma_1 \, \xi^{2q_1}) }{ (1- \gamma_1 \, \xi^{2q_1})} \,,  \\ 
e^{2\mu_2}-1    ~=~& -(\param1-2) \, \frac{\gamma_1 }{ 1 + \gamma_1}\, \frac{1- \, \xi^{2q_1} }{1- \gamma_1 \, \xi^{2q_1}}  \,,     
\end{aligned}
    \label{Layer1bsol4}
\end{equation}
\begin{equation}
\frac{k}{1-\xi^2}   ~=~   \frac{q_2 \, \xi^{2q_2}}{1-  \, \xi^{2q_2}} ~-~  \frac{ q_1\, \gamma_1 \, \xi^{2q_1}}{1- \gamma_1 \, \xi^{2q_1}}  ~-~  ( \param1 -2)  \,  \frac{(1 -\gamma_1) }{ (1 +\gamma_1)}\, \frac{q_1\,  \gamma_1 \,\xi^{2q_1}}{ (1- \gamma_1 \, \xi^{2q_1})^2}   \,.
    \label{ksol3}
\end{equation}
In particular, this implies
\begin{equation}
\big( e^{2\mu_1} ~-1 \big)   ~+~ \big( e^{2\mu_2} ~-1 \big) ~-~ \coeff{1}{2}\, \param1 \,\big(e^{2\mu_0}  ~-1\big)     ~=~   0    \,.
\label{Amusing1}
\end{equation}
One should  also  recall that there is the constraint (\ref{constraint1}), which now becomes
\begin{equation}
 q_1^2 \gamma_1 ~-~   \beta^2\,  q_2^2   ~=~ 0\,.
  \label{constraint2}
\end{equation}

To summarise how the results from this and the previous section combine. Demanding that the scalars, $\mu_j$, vanish at infinity, we have the same constraints, irrespective of the value of $\sigma$:
\begin{gather}
q_2=1,\quad 0\le\gamma_1\le 1,\quad\gamma_2=1,\quad q_1^2\gamma_1=\beta^2,\quad c_4=-2,\quad c_5=2,\quad c_8=1-q_1,\notag\\
c_6=2,\quad c_7=\frac{(1-\gamma_1)\,\sigma - 4}{1+\gamma_1}.\label{eq:GenConstr}
\end{gather}
On the other hand, if the $\mu$ scalars are not required to vanish at infinity, then we have 2 cases:
\begin{gather}
\sigma\neq0,\quad q_2=1,\quad 0\le\gamma_1\le 1,\quad\gamma_2=1,\quad q_1^2\gamma_1=\beta^2,\quad c_4=-2,\quad c_5=2,\quad c_8=1-q_1,\label{eq:Constr1}
\end{gather}
and
\begin{gather}
\sigma=0,\quad q_2=1,\quad \gamma_1,\gamma_2\leq 1,\quad q_1^2\gamma_1=\gamma_2\,\beta^2,\quad c_4=-2,\quad c_5=2,\quad c_8=1-q_1.\label{eq:Constr2}
\end{gather}
Each of \eqref{eq:GenConstr}, \eqref{eq:Constr1} and \eqref{eq:Constr2} leads to all the rest of the functions, given in Appendix \ref{app:restSols}, being regular as well. 

To match with the solution given in  \cite{Ganchev:2021ewa} and Section \ref{sub:a_special_locus}, one must take \eqref{eq:GenConstr} and
\begin{equation}
\sigma ~=~ 2\,, \qquad q_1 ~=~ (2p+1)\,, \qquad   \beta ~=~ - (2p+1) \, \gamma^2\,, \qquad  \gamma_1~=~ \gamma^4 \,. 
\label{reduction2}
\end{equation}
The standard $(1,0,n)$ superstratum solution, given in Section \ref{ss:superstratum}, is obtained in the same case, \eqref{eq:GenConstr}, by setting $q_1=2 n+1$, $q_2=1$, $\gamma_1 = 0$ and taking the limit $\sigma\to\infty$, $\beta \to 0$, with $\param1 \beta = - \frac{1}{4} \alpha^2$ finite.

\section{Features of the six-dimensional uplift}
\label{sec:uplift}
 
The complete uplift formulae are given in \cite{Mayerson:2020tcl}, however we will focus on a subset of the fields, and especially upon the metric.   
To write the Ansatz for the uplift one defines the $S^3$ by introducing four Cartesian coordinates, $x^I$, on $\IR^4$,  obeying  the constraint $x^I x^I=1$. One also defines 
\begin{equation}
 \Delta ~=~  m_{IJ}x^I x^J \,. 
\label{eq:DeltaDef}
\end{equation}
where $m_{IJ}$ is the matrix of scalars,  (\ref{Mdefn}), in three dimensions.
The six-dimensional metric ansatz is then given by: 
\begin{equation}
 \label{eq:6Dmetricuplift} ds_6^2 ~=~ - (\det m_{IJ})^{-1/2} \Delta^{1/2} \, ds_3^2  ~+~ g_0^{-2}(\det m_{IJ})^{1/2} \Delta^{-1/2} m^{IJ} \cD x^I \cD x^J,
\end{equation}
where $ds_3^2$ is the three-dimensional metric, (\ref{genmet1}), and  $\cD$ is the covariant derivative defined in (\ref{covderiv}). The $-$ sign in front of  $ds_3^2 $ is to convert the metric into ``mostly plus'' signature.

The six-dimensional scalars, the dilaton and axion, are given by:
\begin{equation}
 e^{-\sqrt{2}\varphi} ~=~  \Delta, \qquad X  ~=~    \chi_I x^I  \,.
  \label{eq:6Dscalaransatz}
\end{equation}

The  standard BPS form of the six-dimensional metric is:
\begin{equation}
ds_6^2 ~=~ -\frac{2}{\sqrt{\cP}} \, (dv+\hat\beta) \big(du +  \omega + \tfrac{1}{2}\, \mathcal{F} \, (dv+\hat\beta)\big) 
~+~  \sqrt{\cP} \, ds_4^2\,, 
 \label{sixmet}
\end{equation}
where $ds_4^2$ is a metric on a base manifold, $\cB$.    (We have relabelled the fibration vector as $\hat \beta$ so as to avoid confusion with the parameter, $\beta$, in our solutions.) Our primary goal is to use (\ref{eq:6Dmetricuplift}) to determine the various pieces of this metric.  Indeed, 
supersymmetry requires the base metric, $ds_4^2$, to be hyper-K\"ahler and $d \hat\beta$ to be self dual on $\cB$.  As we will see, the result is a new family of elliptically deformed, two centered ambi-polar hyper-K\"ahler metrics. 

We should note, at this point, that one might have expected to get a standard superstratum in which the base metric is $\IR^4$ and the fields depend on the $v$-coordinate in six dimensions.   However, at the outset, we chose a gauge in which the fields are independent of $\psi$, and hence $v$. This gauge choice is equivalent to a spectral flow that converts the canonical $\IR^4$ base into a two centered ambi-polar hyper-K\"ahler metric.

While we are not going to compute the detailed uplifts of the tensor gauge fields, we note that the tensor gauge field components, usually denoted by $\Theta^{(4)}$, are proportional to the components, $B^4_{ij}$, in the uplift formula given in \cite{Mayerson:2020tcl}:
\begin{equation} 
B^4_{ij}  ~=~ \left(-\frac{\sqrt{2}}{g_0^2}\right) \, \frac12\,  \mathring{\omega}_{ijk}\,  \mathring{g}^{kl}\,  \Delta^{1/2}\, \partial_l \left(\Delta^{-1/2} X \right)\,.
\label{eq:6DansatzB4ij}  
\end{equation}
Note that the scalar, $X$, is the axion, whose uplift is given in   (\ref{eq:6Dscalaransatz}), and this is the field that one calls $Z_4$ in six dimensions.

The important thing to note here is that if $\param1 =0$, then $\nu \equiv 0$, which means that  $X \equiv 0$ and hence  $(Z_4,\Theta^{(4)})$ vanish identically.   In the standard superstratum, the pair, $(Z_4,\Theta^{(4)})$, are the momentum-carrying ``seeds'' of the solution.  However, in our work, the field $\mu_0$ can also carry momentum charge and in Section \ref{sec:Example} we will see how this can create a scaling superstratum.

\subsection{The four-dimensional base metric}
\label{sec:base}

A straightforward but somewhat involved computation leads to the metric, $ds_4^2$, on the base manifold, $\cB$.  One finds that this metric only depends on the four functions,
$\mu_0$, $F$, $G$ and $H$ that appear in the lowest layer of the BPS equations described in Section \ref{ss:BPSlayers}.
We parametrize the $S^3 \subset \IR^4$ using polar coordinates by taking:
\begin{equation} 
x^1 ~=~ \sin \theta \, \sin \varphi  \,,\qquad x^2 ~=~  \sin \theta \, \cos \varphi \,,\qquad x^3 ~=~  \cos \theta \, \sin \chi \,,\qquad x^4 ~=~  \cos \theta \, \cos \chi   \,.
\label{eq:cartesians}  
\end{equation}

As we will describe in detail below, we obtain a new family of ambi-polar, hyper-K\"ahler geometries, parametrized by $\gamma_1$, $\gamma_2$, $q_1$ and $q_2$.  Indeed, the first-order system, (\ref{BaseSystem}), is precisely what defines the hyper-K\"ahler structure.  
We will also see that, for  $\gamma_2 =1$, these geometries are asymptotically locally-Euclidean (ALE)  \cite{Gibbons:1978tef} in that they are asymptotic to the flat metric on $\IR^4/\ZZ_{q_2}$ at infinity ($\xi \to 1$).   Thus, the  choice $q_2=1$ makes the  base geometries asymptotically Euclidean, and the coordinate reparametrization $\xi \to \xi^\lambda$ used in Section \ref{ss:basesystem} to set $q_2=1$ may be viewed as ``undoing'' the $\ZZ_{q_2}$ orbifold.  The ambi-polar structure of the metric then emerges  through individual geometric charges, $\frac{1}{2} (q_2 + q_1)$  and  $\frac{1}{2} (q_2 - q_1)$, in the interior.      

It should be stressed that these geometries are not the canonical Gibbons-Hawking ALE metrics because they only have one $U(1)$ isometry and, as we will see in Section \ref{ss:cplxstr}, this isometry is {\it not} tri-holomorphic. These metrics are expressed in terms of rational functions of $\xi$, and are therefore not some variant of the Atiyah-Hitchin metric.  They are thus  completely new ambi-polar, hyper-K\"ahler geometries.

\subsubsection{The base metric}
\label{ss:basemetric}

The metric, $ds_4^2$, may be written as the sum of the squares of the frames:
\begin{equation}
ds_4^2= a^2 \sum_{i=1}^{4}\big(e^i\big)^2,
\end{equation}
\begin{equation}
\begin{aligned}
e^1  ~=~ & \sqrt{\frac{H(\xi)\,  \Gamma}{2} }  \, \frac{d \xi }{\xi} \,,  \\
e^2  ~=~ &  \sqrt{\frac{H(\xi)}{2\, \Gamma} } \, \Big[\sinh (2\mu_0) \,  \sin \theta \cos \theta \sin 2\varphi \,d \theta    \\
& \qquad \qquad  \qquad~- \big(\cosh (2\mu_0) - \sinh (2\mu_0)\, \cos 2\varphi \,  \big) \sin^2 \theta \, d\varphi~+~ \cos^2 \theta\, d \chi \, \Big]\,,  \\
e^3  ~=~ &  \frac{1}{\sqrt{\Gamma} } \, \Big[ \big(F(\xi) \, e^{-\mu_0} \sin^2 \theta  ~-~  G(\xi) \, e^{\mu_0} \cos^2 \theta\,\big)  \,   \cos \varphi \,d \theta    \\
& \qquad \qquad  \qquad \qquad  \qquad~-  e^{\mu_0}    \sin \theta \cos \theta \, \sin \varphi \, \big( F(\xi) \, d \chi  -  G(\xi) \, d \varphi  \,\big)  \, \Big]\,, \\
e^4  ~=~ &  \frac{1}{\sqrt{\Gamma} } \, \Big[ \big(F(\xi) \, e^{\mu_0} \sin^2 \theta  ~-~ G(\xi) \, e^{-\mu_0} \cos^2 \theta\,\big)  \,   \sin \varphi \,d \theta    \\
& \qquad \qquad  \qquad \qquad  \qquad~+ e^{-\mu_0}    \sin \theta \cos \theta \, \cos \varphi \, \big( F(\xi) \, d \chi  -  G(\xi) \, d \varphi  \,\big)  \, \Big]  \,,
\end{aligned}
 \label{baseframes}
\end{equation}
where
\begin{equation}
 \Gamma  ~\equiv~ G(\xi) \, \cos^2 \theta ~-~ F(\xi) \, \big(\cosh (2\mu_0) - \sinh (2\mu_0)\, \cos 2\varphi \,  \big)\, \sin^2 \theta  \,.
 \label{Gammadefn}
\end{equation}

This metric is positive definite for $q_1 =q_2$ and ambi-polar for $q_1>q_2$.  The latter is easily seen by taking  $\xi \to 0$, where $\mu_0 \to 0$ and $F \to \frac{1}{2} (q_1-q_2)$,  $G \to \frac{1}{2} (q_1+q_2)$, whence:
\begin{equation}
 \lim\limits_{\xi\rightarrow 0}\Gamma  ~=~ \frac{1}{2}  \, \big(\, q_2 ~+~ q_1 \cos 2\theta \,  \big)\,.
 \label{Gammalim1}
\end{equation}
Since one has $0 < \theta < \frac{\pi}{2}$, this changes sign for $ q_1 > q_2$.  The general expression for $\mu_0 =0$ is:
\begin{equation}
 \Gamma  ~=~ \coeff{1}{2}\,( G(\xi)-F(\xi))  ~+~  \coeff{1}{2}\,( G(\xi)+F(\xi))  \, \cos2 \theta  ~=~ \frac{q_2}{2} \,\frac{(1+ \gamma_2 \, \xi^{2q_2})}{(1- \gamma_2 \, \xi^{2q_2})}  ~+~   \frac{q_1}{2} \,\frac{(1+ \gamma_1\, \xi^{2q_1})}{(1- \gamma_1 \, \xi^{2q_1})} \, \cos2 \theta    \,.
 \label{Gammared}
\end{equation}

We have computed the curvature of $ds_4^2$ and it is, as one should expect, non-trivial and self-dual for $q_1>q_2$ and that of flat $\IR^4$ for $q_1=q_2$.

\subsubsection{The complex structures}
\label{ss:cplxstr}

There is an obvious candidate for a K\"ahler form:
\begin{equation}
J   ~\equiv~  e^1 \wedge  e^2 ~-~ e^3 \wedge  e^4 \,.
 \label{Jform}
\end{equation}
This is manifestly anti-self-dual, and one can easily verify that it is closed as a result of the differential identities (\ref{BaseSystem}).  The other two complex structures, $K$ and $L$,  are almost as simple, and are defined by: 
\begin{equation}
\begin{aligned}
K  ~\equiv~&  (e^1 \wedge  e^3 ~+~ e^2\wedge  e^4)\, \cos \chi ~+~   (e^1 \wedge  e^4 ~-~ e^2\wedge  e^3)\, \sin \chi \,, \\ 
L  ~\equiv~&  -(e^1 \wedge  e^3 ~+~ e^2\wedge  e^4)\, \sin \chi  ~+~  (e^1 \wedge  e^4 ~-~ e^2\wedge  e^3)  \, \cos \chi  \,.
\end{aligned}
 \label{KLform}
\end{equation}
These are again manifestly anti-self-dual and are closed by virtue of the differential identities (\ref{BaseSystem}).

One should note that  $\frac{\partial}{\partial \chi}$ is a Killing vector of the metric $ds_4^2$, but it is not tri-holomorphic because of the $\chi$-dependence in  (\ref{KLform}).

It is elementary to obtain potentials for these K\"ahler forms:
\begin{equation}
\begin{aligned}
J  ~=~& d A \,, \qquad K + i \, L   ~=~  d B  \,, \\
A ~=~ & - \frac{1}{2}\,\bigg[\, \frac{1}{2\, \xi}\, H(\xi) \,\sinh (2 \mu_0) \, \sin^2 \theta\,\sin 2 \varphi \, d \xi  ~+~  G(\xi) \, \sin^2 \theta \, d \varphi ~+~F(\xi) \, \cos^2 \theta \, d \chi  \,\bigg], \\ 
B~=~ & e^{-i\, \chi} \, \big( e^{\mu_0} \cos \varphi  + i\,  e^{-\mu_0}\sin  \varphi \big) \sqrt{ \frac{H(\xi)}{2}}\, \, \bigg[\, \frac{G(\xi)}{\xi}\,   \sin  \theta\, \cos  \theta\,d \xi  -  \sin^2 \theta \, d \theta - i\, \sin  \theta\, \cos  \theta\, d \chi  \,\bigg]   \,.
\end{aligned}
 \label{JKLpots}
\end{equation}

While one can choose any linear combination of $J,K$ and $L$ to define a complex structure, we will use ${J^\mu}_\nu$ as the complex structure henceforth.

\subsubsection{Some complex coordinates and the hermitian form of the metric}
\label{ss:cplxcoordmet}

Using  ${J^\mu}_\nu$ as the complex structure, one can define holomorphic coordinates:
\begin{equation}
z_1   ~\equiv~  \frac{\big( e^{\mu_0} \cos \varphi  -  i\,  e^{-\mu_0}\sin  \varphi \big) }{\sqrt{\sinh 2 \mu_0}}\, \sin \theta \,, \qquad z_2   ~\equiv~    e^{i\, \chi} \, \sqrt{H(\xi) \,\sinh 2 \mu_0}  \, \cos \theta   \,.
 \label{cplxcoords}
\end{equation}
The metric, $ds_4^2$ can then be written in hermitian form:
\begin{equation}
ds_4^2 ~=~ A_1 \, \big| d z_1\big|^2 ~+~ A_2 \, \big| d z_2 \big|^2 ~+~ A_3 \, \bigg( \frac{d z_1}{z_1} \,  \frac{d \bar z_2}{\bar z_2}  +   \frac{d \bar z_1}{\bar z_1} \,    \frac{d z_2}{z_2} \bigg) ~-~ i\,A_4\, \bigg( \frac{d z_1}{z_1} \,  \frac{d \bar z_2}{\bar z_2}  -   \frac{d \bar z_1}{\bar z_1} \,    \frac{d z_2}{z_2} \bigg) \,.
 \label{herm-met}
\end{equation}
We then find:
\begin{equation}
\begin{aligned}
A_1~=~& 2\, 
 \sinh (2\mu_0) \, \Big[  G(\xi) ~+~   \cQ \,\big(\cosh (2\mu_0) - \sinh (2\mu_0)\, \cos 2\varphi \,  \big)\, \sin^2 \theta \, \Big] \,, \\
A_2~=~&  \frac{2}{ H(\xi)\,\sinh (2\mu_0)}\,    \big[  -F (\xi) ~+~   \cQ \, \cos^2 \theta \,\big]    \,, \\
A_3~=~&   \cQ \, \sin^2 \theta  \, \cos^2 \theta  \,, \qquad
A_4~=~ - \cQ \,  \sinh (2\mu_0)\,  \sin^2 \theta  \, \cos^2 \theta \,  \sin 2\varphi     \,.
\end{aligned}
 \label{Acoeffs}
\end{equation}
where
\begin{equation}
\cQ   ~\equiv~   \frac{1 }{ \Gamma} \,\big( F(\xi) \, G(\xi) + \coeff{1}{2} \,H(\xi)\big) ~=~ \frac{1}{ 4\,\Gamma} \,(q_1^2 -q_2^2)  \,.
 \label{cQdefn}
\end{equation}

It is interesting to note that for $q_1  = q_2$ one has $\cQ  \equiv 0$, which means  $H(\xi) = -2 F(\xi) \, G(\xi) $.  Moreover, one has:
\begin{equation}
G \,\sinh (2\mu_0) ~=~    \frac{2\, q_1\, \beta}{1- \beta^2}\,.
 \label{Gsinhsimp}
\end{equation}
As a result, the metric defined by  (\ref{herm-met}) and  (\ref{Acoeffs}) becomes

\begin{equation}
ds_4^2 ~=~ A_1 \, \big| d z_1\big|^2 ~+~ A_2 \, \big| d z_2 \big|^2 ~=~ \frac{4\,q_1 \beta}{(1- \beta^2)}\, \big| d z_1\big|^2 ~+~\frac{(1- \beta^2)}{2\, q_1 \beta}\, \, \big| d z_2 \big|^2   \,.
 \label{herm-met-simp}
\end{equation}
This is manifestly a flat metric in (scaled) Cartesian coordinates.

\subsubsection{A K\"ahler potential}
\label{ss:KahlerPot}

One can write the standard expression for the K\"ahler potential, $\cK$, in terms of real variables implicitly as  
\begin{equation}
A_\mu   ~=~   \partial_\mu \, \cU ~+~ {J_\mu}^\nu\,  \partial_\nu\, \cK  \,.
 \label{Kpotdefn1}
\end{equation}
where the potential, $A_\mu$, is given in (\ref{JKLpots}) and $J$ is the complex structure defined by (\ref{Jform}).  The function, $\cU$, is simply a gauge transformation reflecting an implicit gauge choice in (\ref{JKLpots}). 

A straightforward computation leads to a solution:
\begin{equation}
\begin{aligned}
\cU ~=~& \frac{\beta}{8}\,(q_1+q_2) \, \frac{ \xi^{q_1-q_2}(1-  \gamma_2 \, \xi^{2q_2})}{(1- \gamma_1 \, \xi^{2q_1})} ~-~ \frac{1}{4}\,\bigg( \frac{q_1 }{\beta\,\xi^{q_1-q_2} (1-  \gamma_2 \, \xi^{2q_2})} ~+~  \frac{q_2\, \beta\,\xi^{q_1-q_2}  }{(1- \gamma_1 \, \xi^{2q_1})}\bigg) \,, \\
\cK ~=~ & \frac{1}{4}\,\bigg[ \frac{q_1 }{(1- \gamma_1 \, \xi^{2q_1})} ~-~   \frac{q_2}{(1- \gamma_2 \, \xi^{2q_2})}  ~+~ \coeff{1}{2}\,(q_1^2 - q_2^2)\, \log \xi ~-~ q_1\, \cos^2 \theta \\
& \quad +\bigg(  \frac{q_1 }{\beta\,\xi^{q_1-q_2} (1-  \gamma_2 \, \xi^{2q_2})} ~-~ \frac{q_2\, \beta\,\xi^{q_1-q_2}  }{(1- \gamma_1 \, \xi^{2q_1})} \\
& \qquad\qquad\qquad\qquad\qquad\qquad~+~  \frac{\beta}{2}\,(q_1+q_2) \,\xi^{q_1-q_2}\, \frac{ (1-  \gamma_2 \, \xi^{2q_2})}{(1- \gamma_1 \, \xi^{2q_1})} \bigg)\sin^2 \theta \cos 2 \varphi  \bigg]     \,.
\end{aligned}
 \label{Kpot1}
\end{equation}

\subsection{The other parts of the six-dimensional metric}
\label{sec:6Drest}

\subsubsection{The warp factor}
\label{ss:warp}

The warp factor, $\cP$, is relatively straightforward.  One finds that  
\begin{equation}
\cP   ~=~  \frac{ \Delta }{g_0^4\, a^4 \, \Gamma^2}  \,,
 \label{Pfactor}
\end{equation}
where
\begin{equation}
\Delta   ~\equiv~  m_{AB} x^A x^B ~=~e^{2\, \mu_2} \cos^2 \theta ~+~ e^{2\, \mu_1}  \big(\cosh (2\mu_0) - \sinh (2\mu_0)\, \cos 2\varphi \,  \big)\, \sin^2 \theta \,.
 \label{Deltadefn}
\end{equation}
%

\subsubsection{The fibration vector}
\label{ss:fibration}

The connection, $\hat\beta$, of the $v$-fibration is given by:
\begin{equation}
\begin{aligned}
\hat\beta  ~=~ &   -\frac{ R_y  }{\sqrt{H(\xi)\, \Gamma} }\, e^2  \\
~=~ & -\frac{ R_y  }{\sqrt{2}\, \Gamma} \,  \Big[ \sinh (2\mu_0) \,  \sin \theta \cos \theta \sin 2\varphi \,d \theta    \\
& \qquad \qquad  \qquad~- \big(\cosh (2\mu_0) - \sinh (2\mu_0)\, \cos 2\varphi \,  \big) \sin^2 \theta \, d\varphi~+~ \cos^2 \theta\, d \chi \, \Big]\,.
 \end{aligned}
 \label{betavec}
\end{equation}
where $e^2$ is one of the frames in (\ref{baseframes}).

One can then verify that one has:
\begin{equation}
\begin{aligned}
\Theta_3  ~=~ &  d \hat\beta  \\
~=~ & \frac{\sqrt{2}\, R_y  }{ \Gamma^2} \, \big(\cosh (2\mu_0) - \sinh (2\mu_0)\, \sin^2 \theta \cos 2\varphi \,  \big)(e^1 \wedge  e^2 ~+~ e^3 \wedge  e^4) \\
&+ \frac{2\, R_y  \, F(\xi) }{  \sqrt{H(\xi)}\,  \Gamma^2}\, \sinh (2\mu_0)   \,  \sin \theta \cos \theta \ \Big( e^{\mu_0} \sin \varphi\, (e^1 \wedge  e^3 ~-~ e^2\wedge  e^4) \\
& \qquad \qquad\qquad\qquad\qquad\qquad\qquad\qquad+~  e^{-\mu_0} \cos \varphi\, (e^1 \wedge  e^4 ~+~ e^2\wedge  e^3)\Big) \Big]\,.
 \end{aligned}
 \label{Theta3}
\end{equation}
This is manifestly a self-dual flux.

\subsubsection{The momentum function and angular momentum vector}
\label{ss:momparts}

The momentum function, $\cF$, is given by:
\begin{equation}
\begin{aligned}
\cF ~=~  2 -\frac{ 2}{R_y^2 a^2 g_0^4 }  \,  \bigg[  &  \frac{1}{\Gamma}\,\big(e^{2 \mu_2} \, F(\xi)^2  +  \coeff{1}{2}\, e^{2 \mu_1} \,  H(\xi)\big )  \big(\cosh (2\mu_0) - \sinh (2\mu_0)\, \cos 2\varphi \,  \big)  \sin^2 \theta \\
& ~+ \frac{1}{\Gamma}\,\big(e^{2 \mu_1} \, G(\xi)^2  +  \coeff{1}{2}\, e^{2 \mu_2} \,  H(\xi)\big )    \cos^2 \theta  ~-~ \frac{2 k }{1 - \xi^2}\, \bigg]\,.
 \end{aligned}
 \label{Fres}
\end{equation}

The non-trivial components of the angular momentum vector are:
\begin{equation}
\omega  ~\equiv~  \omega_\theta \, d \theta ~+~ \omega_\varphi \, d \varphi ~+~ \omega_\chi \, d \chi  \,,
 \label{omvec}
\end{equation}
with
\begin{equation}
\begin{aligned}
\omega_\theta  ~=~&  -\frac{ \sqrt{2} }{R_y\, a^2 g_0^4 \, \Gamma}\,  \sinh (2\mu_0)\,\sin\theta  \cos \theta \, \sin 2\varphi  \,\, \qty(e^{2\mu_1} G(\xi) + \omega_0)    \,, \\ 
\omega_\varphi  ~=~ &  \frac{ \sqrt{2} }{R_y\, a^2 g_0^4\, \Gamma } \,\big(\cosh (2\mu_0) - \sinh (2\mu_0)\, \cos 2\varphi \,  \big) \, \sin^2\theta \, \qty(e^{2\mu_1} G(\xi) + \omega_0)   \,,\\
\omega_\chi   ~=~ &  \frac{ \sqrt{2} }{R_y\, a^2 g_0^4\, \Gamma} \, \cos^2\theta    \, \qty(e^{2\mu_2} F(\xi) - \omega_0)  
 \end{aligned}
 \label{omres}
\end{equation}
where $\cQ$ is defined in (\ref{cQdefn}) and
\begin{equation}
\begin{aligned}
\omega_0  ~\equiv~& - \frac{ k }{1 - \xi^2} -  \frac{1}{2}\,R_y^2 \,a^2 g_0^4  \, +\cQ\,\Delta   \,.
 \end{aligned}
 \label{om0defn}
\end{equation}

%

\subsection{Asymptotics and conserved charges}

We can extract the conserved charges of the geometries according to the methods of \cite{Bena:2015bea}.

At large $r$, the momentum function limits to a constant value:
\begin{equation}
    \lim_{r\to\infty} \cF ~=~ 2 \,+\, \frac{2}{ R_y^2\,a^2 g_0^4} \qty(2 q_1 - q_2 - \frac{q_1 \sigma}{1 - \sqrt{\gamma_1}} + \frac{q_1 (\sigma -2)}{1 + \gamma_1}) \,.
\end{equation}

In order to get the correct $AdS_3 \times S^3$ asymptotics, this constant must vanish \cite{Bena:2015bea,Mayerson:2020tcl}. This leads to the following constraint, which is akin to what is often called the ``regularity condition'' of superstrata:
\begin{equation}
    R_y^2\,a^2 g_0^4 ~=~ q_2 - 2 q_1 + \frac{q_1 \param1}{1 - \sqrt{\gamma_1}} - \frac{q_1 (\param1 -2)}{1 + \gamma_1} \,.
    \label{constraint_a}
\end{equation}
As we will discuss below, the positivity of the left-hand side places bounds on $\param1$ and $\gamma_1$.

Upon applying the constraint, the momentum function then decays at infinity as
\begin{equation}
    \cF ~=~ - \frac{q_1^2 - q_2^2}{2 R_y^2 \,g_0^4 \, r^2} + O\qty(\frac ar)^4 \,.
\end{equation}

The fibration and angular momentum vectors, $\hat\beta$ and $\omega$, also decreases as $1/r^2$ at infinity:

\begin{equation}
\begin{aligned}
    \hat\beta ~&=~ \frac{R_y}{\sqrt{2}} \qty(\sin^2\theta \, d\varphi - \cos^2\theta \, d\chi) \frac{a^2}{r^2} \,+\,O\qty(\frac ar)^4 \,,
    \\
    \omega ~&=~ \frac{R_y}{\sqrt{2}} \qty[1+ \frac{q_1-q_2}{R_y^2 \, a^2 g_0^4}] \qty(\sin^2\theta \, d\varphi + \cos^2\theta \, d\chi) \frac{a^2}{r^2} \,+\, O\qty(\frac ar)^4\,. 
\end{aligned}
\end{equation}

Using these results, one can then follow the procedure described in \cite{Bena:2015bea} to read-off the conserved charges\footnote{We are using the  standard conventions in which the excitations of the supertube are in the left-moving sector.   Section D.5 of \cite{Mayerson:2020tcl} uses the opposition conventions, swapping $J_L$ and $J_R$.}:
\begin{equation}
\begin{aligned}
    \hat\beta_\phi + \hat\beta_\chi + \omega_\phi + \omega_\chi ~&=~ \frac{\sqrt{2}}{r^2}(J_L - J_R \cos2\theta) + \mathcal{O}(r^{-4}) \,,\\ \cF ~&=~ - \frac{2Q_P}{r^2} +\mathcal{O}(r^{-4}) \,,
\end{aligned}
\end{equation}
leading to:
\begin{equation}
    J_L ~=~ \frac{R_y}{2} \qty(a^2 + \frac{q_1- q_2}{R_y^2 g_0^4})\,, \qquad 
    J_R ~=~ \frac{a^2 R_y}2 \,, \qquad 
Q_P ~=~ \frac{q_1^2 - q_2^2}{4 R_y^2 \,g_0^4} \,.
  \label{SScharges}
\end{equation}


\subsection{Inverting the spectral flow}

Because of the gauge choices described in Section \ref{sub:qball}, our six-dimensional solution is $v$-independent even though it carries momentum charge.  The more canonical formulation as a superstratum would involve a topologically trivial $\IR^4$ base and would have  explicit $v$-dependence in the fluctuating modes, and trivial constant terms in the magnetic potentials.   To convert the canonical superstratum into the solution presented here one needs to do a spectral flow of the form:
\begin{equation}
 \varphi~\to~ \varphi ~-~ \coeff{1}{2}\,(q_1 -q_2) \, \psi \,,  \qquad  \chi~\to~ \chi ~-~ \coeff{1}{2}\,(q_1 -q_2) \,  \psi  \,,
 \label{spec-flow1}
\end{equation}
which means the Hopf fiber of the Gibbons-Hawking space is shifted by:
\begin{equation}
 (\varphi+\chi)  ~\to~  (\varphi+\chi)  ~-~ (q_1 -q_2) \, \psi  \,.
 \label{spec-flow2}
\end{equation}
This flow takes the trivial base to a base with GH charges $\frac{1}{2} (q_1 + q_2)$ and $\frac{1}{2} (q_1 -q_2)$.  It also shifts the quantized CFT charges according to \cite{Balasubramanian:2000rt,Maldacena:2000dr,Mathur:2001pz,Lunin:2004uu,Bena:2008wt}:
\begin{equation}
j_L  ~\to~ j_L ~+~ \frac{c}{12}  (q_1 -q_2) \,, \qquad h   ~\to~h   ~+~  (q_1 -q_2) j_L  ~+~ \frac{c}{24}  (q_1 -q_2)^2   \,.
 \label{spec-flow3}
\end{equation}
where $c = 6 N_1 N_5$ is the central charge of the D1-D5 CFT. This translates into the following shift of the supergravity charges\footnote{The translation between quantized and supergravity charges can be found in many places, see, for example, \cite{Bena:2015bea,Bena:2016ypk}. }:
\begin{equation}
    J_L ~\to~ J_L + \frac{q_1 - q_2}{2 R_y g_0^4} \,, \qquad
    Q_P ~\to~ Q_P + (q_1 - q_2) \frac{J_L}{R_y} + \frac{(q_1 - q_2)^2}{4 \, R_y^2\, g_0^4}
          \,.
 \label{spec-flow4}
\end{equation}

This means that the canonical form of our new superstratum solution on an $\IR^4$ base with $v$-dependent modes must have
\begin{equation}
    J_L ~=~     J_R ~=~   \frac{a^2 R_y}2 \,, \qquad  Q_P ~= \frac{q_1 - q_2}{2 R_y^2 g_0^4} \qty(q_2 - a^2 R_y^2 g_0^4) \,.
     \label{canonSScharges}
\end{equation}
These charges are determined by the fact that the spectral flow (\ref{spec-flow4}) applied to them results in the charges  (\ref{SScharges}).
The canonical superstratum therefore has $J_L =   J_R  \sim a^2$.  The momentum charge determines the AdS$_3$ radius, $g_0^{-1} \equiv (Q_1 Q_5)^{1/4}$, and is related to the mode numbers and the  amplitudes  through  the constraint (\ref{constraint_a}).  It is interesting to note that the expression for $Q_P$ may be rewritten as:
\begin{equation}
\frac{2\, Q_P}{(q_1 - q_2)}~+~a^2  ~=~  \frac{q_2}{R_y^2 g_0^4}    ~=~  \frac{q_2\, Q_1 Q_5}{R_y^2 }    \,,
     \label{Regcond1}
\end{equation}
which is the analogue of the usual superstratum regularity condition.

\section{An Important Example} 
\label{sec:Example}

The solution with $\param1 =0$ may seem a rather degenerate limit of the solutions we have presented here, but it is perhaps one of the more significant results in this paper. 
As we noted at the beginning of Section  \ref{sec:uplift}, setting $\param1 =0$ removes the standard momentum carriers of the original superstratum.  On the other hand, this solution can still have momentum charge carried by $\mu_0$.  We will therefore examine this solution in more detail.

What makes it important is that the six-dimensional uplift only has excitations in the metric, the gauge fields $(Z_1,\Theta^{(2)})$ and $(Z_2,\Theta^{(1)})$, and the dilaton.  The axion and flux, $(Z_4,\Theta^{(4)})$, are identically zero.  If one takes the S-dual of such a D1-D5 superstratum, then it becomes a {\it pure  F1-NS5 superstratum whose excitations lie only in the NS-NS sector of the theory}:  there are no R-R excitations. This means one has the possibility of using exact world-sheet methods for exploring these geometries.

While such solutions exist for arbitrary  $\gamma_1, \gamma_2 \leq 1$, we will focus on smooth solutions that limit to the supersymmetric critical point at infinity.  That is, we impose (\ref{eq:GenConstr}), with $\sigma=0$, and use the general solution given in \eqref{Omega1constraint}, \eqref{mu0res12}, \eqref{Layer1bsol2}, \eqref{ksol1} and Appendix \ref{app:restSols}: 
\begin{equation}
\begin{aligned}
\nu  ~\equiv~& 0   \,, \qquad e^{2\mu_0} ~=~   \frac{ (1-  \lambda^2 \, \xi^{2q_1}) ~+~ \lambda\, q_1\, \xi^{q_1-1} \,  (1-  \xi^{2})  }{ (1-  \lambda^2  \, \xi^{2q_1}) ~-~ \lambda\, q_1\,  \xi^{q_1-1} \,  (1-  \xi^{2}) }      \,, \\
 e^{2\mu_1}   ~=~& 1~-~  \frac{2 \,\lambda^2 \, (1- \xi^{2q_1}) }{(1 + \lambda^2) (1- \lambda^2 \, \xi^{2q_1})} \,,  \qquad
 e^{2\mu_2}   ~=~ 1~+~  \frac{2 \,\lambda^2 \, (1- \xi^{2q_1}) }{ (1 + \lambda^2)(1- \lambda^2 \, \xi^{2q_1})}  \,,    
\end{aligned}
\label{c3zero-simp}
\end{equation}
where we have solved the constraint in  (\ref{reduction3}) by introducing a new parameter, $\lambda$:
\begin{equation}
\beta  ~\equiv~  \lambda \, q_1\,, \qquad \gamma_1  ~\equiv~  \lambda^2 \,.
 \label{newparams}
\end{equation}

The metric functions\footnote{To go to the canonically normalised time, discussed in the footnote before \eqref{Omega1constraint}, the parameter $s$ of the time rescaling, $\tau\to s\,\tau$, is $s=\frac{1+(1-2\,q_1)\lambda^2}{1+\lambda^2}$.} are: 
\begin{equation}
\Omega_0^2   ~=~ 
 \frac{ (1 - \lambda^2)(1+ \lambda^2 \, \xi^{2q_1}) }{(1 + \lambda^2)(1- \lambda^2 \, \xi^{2q_1})} 
\bigg(1 \,+\,  \frac{2 \,\lambda^2 \, (1- \xi^{2q_1}) }{ (1 + \lambda^2)(1- \lambda^2 \, \xi^{2q_1})} \bigg) \bigg(1 \,-\, \lambda^2 \, q_1^2\, \xi^{2q_1-2}\,\frac{ (1- \xi^2)^2 }{(1- \lambda^2 \, \xi^{2q_1})^2} \bigg)   \,,
  \label{Om0-c3zero}
\end{equation}
\begin{equation}
\Omega_1    ~\equiv~ 1\,, \qquad \frac{k}{1-\xi^2}   ~=~\frac{\xi^2}{1-\xi^2}+\frac{q_1\,\lambda^2\,\xi^{2\,q_1}\Big(1-\lambda^2\big(3-\xi^{2\,q_1}(1+\lambda^2)\big)\Big)}{(1+\lambda^2)(1-\lambda^2\,\xi^{2\,q_1})^2}\,.
  \label{k-c3zero}
\end{equation}

For completeness, we re-state the gauge fields in terms of the other fields and the functions $F$ and $G$, (\ref{FGres1}), with parameters fixed by \eqref{eq:GenConstr}, which provides the easiest way of obtaining expressions for them:
\begin{equation}
\begin{aligned}
  \Phi_1 ~=~&  \frac12 e^{-2\mu_2}  \,, \qquad \Psi_1 ~=~  \frac{1}{2}\, \bigg(F(\xi) ~+~  \frac{k}{ (1-\xi^2) } \, e^{-2\mu_2}  \bigg)  \,,  \\
 \Phi_2 ~=~& \frac12 (1 - e^{-2\mu_1}) \,, \qquad\Psi_2 ~=~  \frac{1}{2}\, \bigg(G(\xi) ~-~ 1 ~-~  \frac{k}{ (1-\xi^2) } \, e^{-2\mu_2}  \bigg)  \,,\\
F   ~=~&  \frac{1}{2} \,\bigg[q_1 \frac{1+ \lambda^2 \, \xi^{2q_1}}{1- \lambda^2 \, \xi^{2q_1}} ~-~   \frac{1+  \, \xi^{2 }}{1-  \, \xi^{2 }} \bigg]    \,, \qquad G   ~=~  \frac{1}{2} \,\bigg[ q_1 \frac{1+ \lambda^2 \, \xi^{2q_1}}{1- \lambda^2 \, \xi^{2q_1}} ~+~   \frac{1+  \, \xi^{2}}{1-  \, \xi^{2 }} \bigg]   \,.
\end{aligned}
    \label{OtherFields}
\end{equation}
%

\section{Asymptotically \texorpdfstring{AdS$_2$ $\times$ S$^1$}{AdS2 x S1} geometries} 
\label{sec:AdS2}


We now return to superstrata with generic values of $\param1$ and to obtain a better understanding these solutions, it is useful to track the evolution of the length of the $\psi$-circle in the 3-dimensional geometry \eqref{genmet1}:
\begin{equation}
    L_\psi(\xi) ~\equiv~ \frac{2 \pi}{R_{AdS}} \sqrt{g_{\psi\psi}}\,.
\end{equation}

We can generally identify three distinct regions:
\begin{itemize}
    \item Close to the origin of space, one has $L_\psi(r) \sim 2 \pi r/a$, the solution caps off smoothly with no conical singularity.

    \item There is an intermediate regime where $L_\psi(\xi)$ remains constant, which means that the size of the circle is fixed, and the geometry is approximately AdS$_2 \times S^1$. We call this region the throat.

    \item Asymptotically, the size of the circle typically grows linearly with $r$, forming an AdS$_3$ region. It is given by
    \begin{equation}
        L_\psi(r) ~\sim_{r\to \infty}~ 2 \pi C \qty(\frac{r}{a}) \,,\quad \text{or equivalently}\quad L_\psi(\xi) ~\sim_{\xi\to 1}~ \frac{2 \pi C}{\sqrt{1-\xi^2}} \,,
    \end{equation}
    where $C$ is a constant that depends on the exact solution we are looking at. We can compute it using \eqref{Hres1} and \eqref{mu0res2}-\eqref{ksol3}. Its value depends on the parameters $\beta$ and $\sigma$, and on the modes $q_1$ and $q_2$:
    \begin{equation}
    C^2 ~=~ q_2 \,-\, 2 q_1 \,+\, \frac{\sigma \,q_1^2}{q_1 - q_2 \,\beta} \,-\, \frac{(\sigma - 2)\, q_1^3}{q_1^2 + q_2^2 \,\beta^2} \,.
    \label{eq:value_c_general}
    \end{equation}
    Note that using the regularity condition \eqref{constraint_a}, one can write $C$ simply in terms of the radii of the supertube locus and of the AdS space:
    \begin{equation}
        C^2 ~=~  R_y^2\, a^2 \, g_0^4 \,.
        \label{link_C_a}
    \end{equation}
\end{itemize}

More precisely, when $C^2$ is positive, $L_\psi$ grows linearly at infinity, and the geometry is asymptotically  AdS$_3$. But at certain values of the parameters, when $C = 0$, the size of the circle is kept finite at infinity, the asymptotically AdS$_3$ region disappears, the throat becomes infinite, and the geometry is asymptotically AdS$_2 \times S^1$. If we keep on increasing the parameters, we reach a regime where $C^2$ is negative, the solution then develops closed time-like curves and is therefore unphysical.

\begin{figure}[ht!]
	\centering
	\includegraphics[width=.5\textwidth]{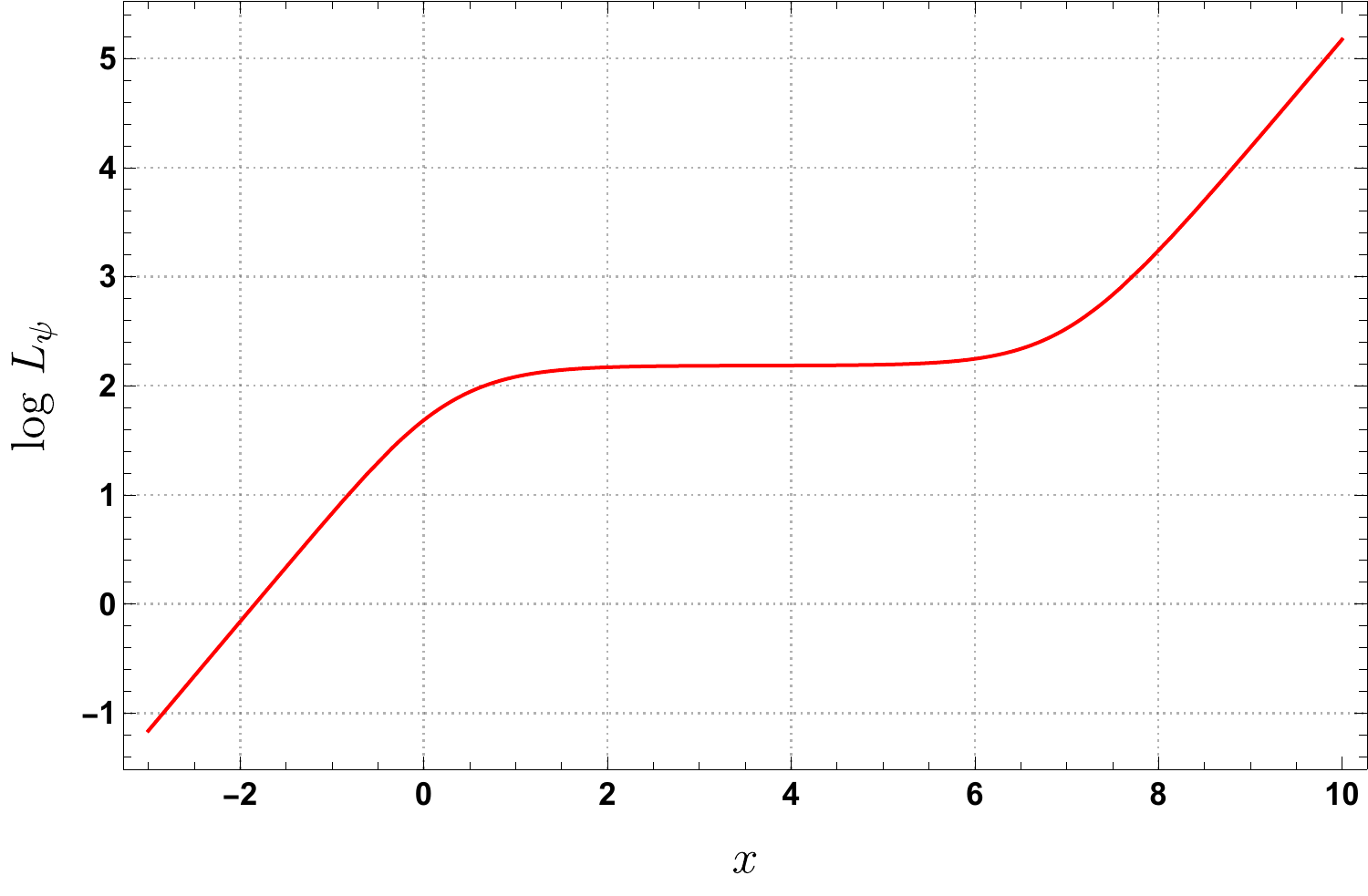}\hspace*{-1mm}
	\includegraphics[width=.51\textwidth]{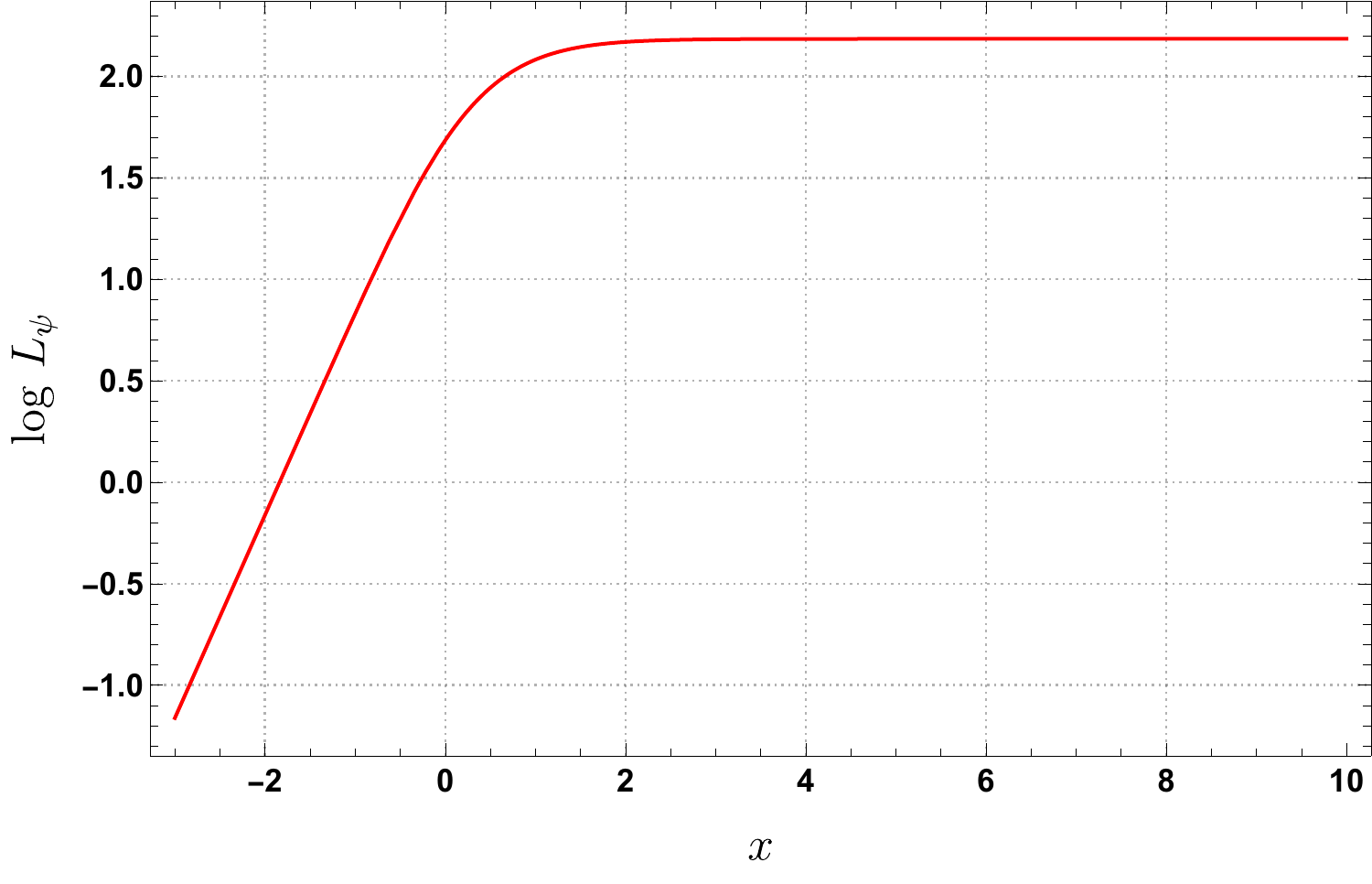}
	\caption{\it Plots of $\log\,L_{\psi}$ against $x$, where $\xi = \frac{e^x}{\sqrt{e^{2x}+1}}$. In the first graph we have taken $\sigma =2$, $\beta = -\frac{q_1}{2q_1 - 1} + \frac{1}{10^6}$, $q_1=5$. Here $C^2$ has a small positive value, and we identify the two $AdS_3$ regions of linear growth separated by a plateau, which is the throat. The second graph corresponds to $\sigma =2$, $\beta = -\frac{q_1}{2q_1 - 1}$, $q_1=5$. For this solution, $C$ is zero, and the asymptotic $AdS_3$ region is not present.}
	\label{fig:gpsipsi12}
\end{figure}

To illustrate this behaviour, we take $q_2 =1$ and plot,  in Fig.~\ref{fig:gpsipsi12}, the logarithm of the length of the circle, $\log\,L_{\psi}$, against $x$, defined through $\xi = \frac{e^x}{\sqrt{e^{2x}+1}}$. This is done, as an example, on the special locus, \eqref{reduction2}. There we obtain
\begin{equation}
C^2 ~=~ \frac{1 - (4\,p+1) \gamma^2}{1 + \gamma^2}=\frac{1 - (2\,q_1-1) \gamma^2}{1 + \gamma^2}\,,
\end{equation}
where $q_1 = 2 p+1$ as in (\ref{reduction2}).
Asking for $C^2$ to be positive, we find again the condition given in \eqref{eq:ctc_limit_special_locus}:
\begin{equation}
\gamma^2\leq\frac{1}{4\,p+1}=\frac{1}{2\,q_1-1}, \quad\text{or equivalently}\quad -\beta\leq\frac{q_1}{2\,q_1-1}.
\end{equation}
A choice of parameters for which this bound is respected is shown on the left graph of Fig.~\ref{fig:gpsipsi12}. The aforementioned three regions are identified as follows: the linear growth on the left corresponds to the cap; the plateau, in the middle, is the throat; and the growing part to the right signifies the AdS$_3$ region. 

\vspace{0.5mm}
\vspace{1mm}
The bound is saturated when $C=0$. The solutions saturating the bound are asymptotically AdS$_2 \times S^1$ geometries, as shown on the right graph of Fig.~\ref{fig:gpsipsi12}.

More generally, moving away from the special locus, one finds that the length of the $\psi$ circle at  infinity when $C=0$ is:
\begin{equation}
L_{\psi}^{(\infty)} ~=~ \sqrt{2 q_2 \,(q_1 - q_2)} \,\pi \,.
\end{equation}

\begin{figure}[ht!]
	\centering
	\includegraphics[width=\textwidth]{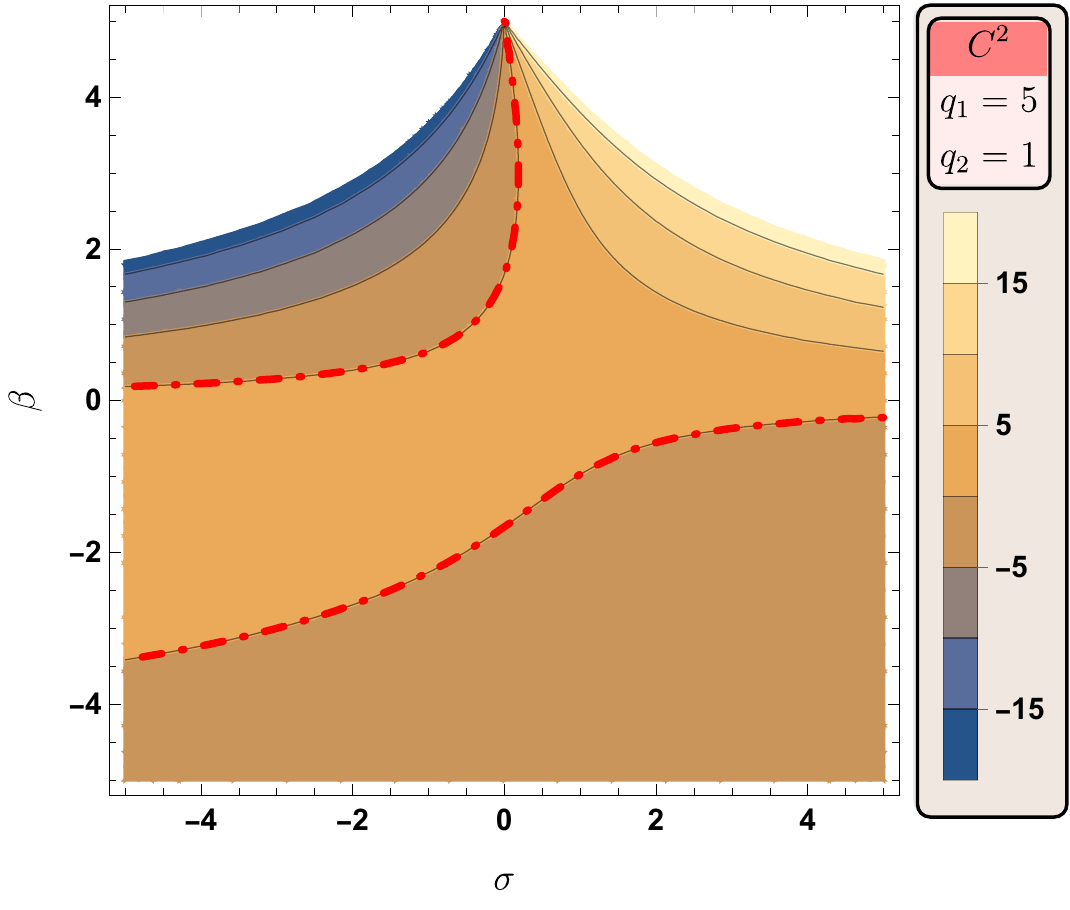}
	\caption{\it Plot of $C^2$, \eqref{eq:value_c_general}, for $q_1=5$, $q_2=1$, as a function of $\sigma$ and $\beta$. The red dashed-dotted lines indicates the $C=0$ contour. In the region between them is where the values are positive. $C^2$ determines the size of the $\psi$ circle at infinity, in the asymptotically AdS$_3$ region.}
	\label{fig:sigma_beta_plane}
\end{figure}

\begin{figure}[ht!]
	\centering
	\includegraphics[width=\textwidth]{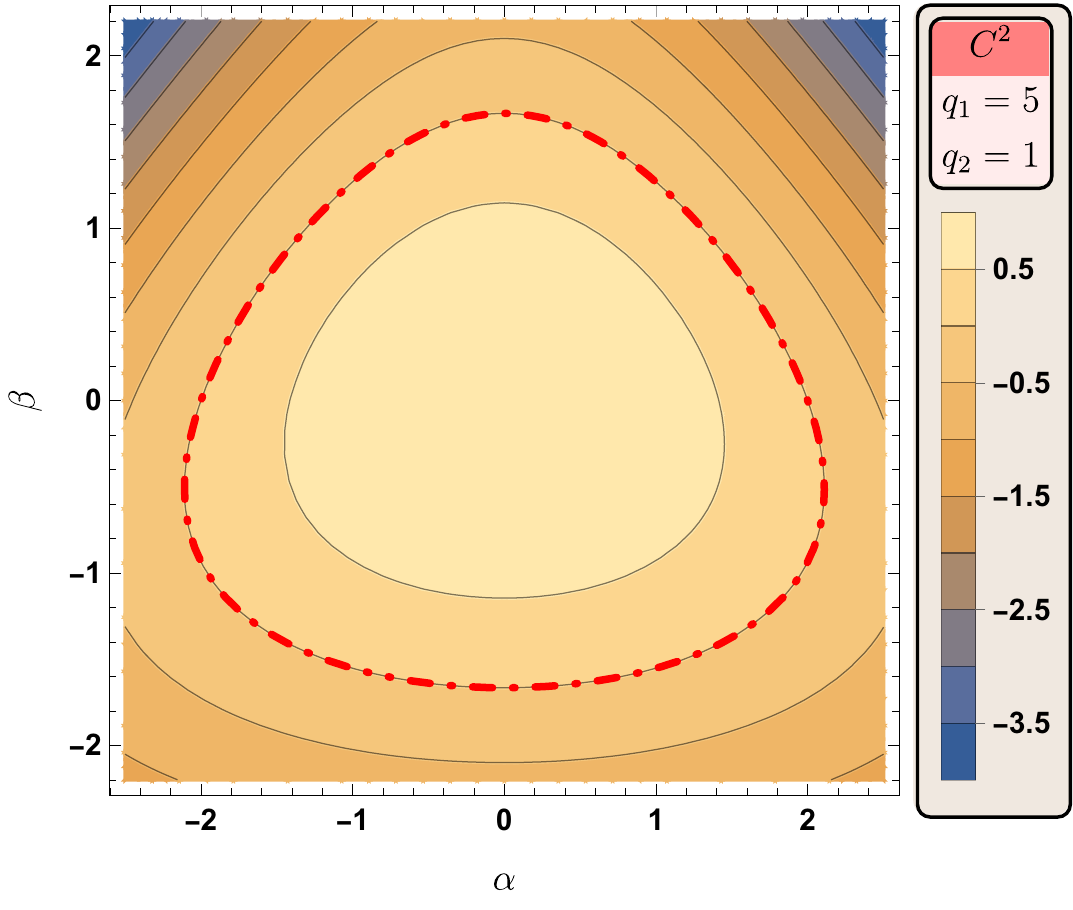}
	\caption{\it Plot of $C^2$, \eqref{eq:value_c_general}, for $q_1=5$, $q_2=1$, as a function of $\alpha$ and $\beta$. The red dashed-dotted lines indicates the $C=0$ contour, which encloses the region of positive values. $C^2$ determines the size of the $\psi$ circle at infinity, in the asymptotically AdS$_3$ region.}
	\label{fig:alpha_beta_plane}
\end{figure}

The solutions of the equation $C = 0$ describe a contour in the plane $(\sigma,\, \beta)$, that separates it into two regions: the physical region, and the region with CTCs. This is represented on Fig.~\ref{fig:sigma_beta_plane}, for $q_1 = 5$ and $q_2 = 1$. Alternatively, one can replace $\sigma$ by $\alpha^2 = - 4 \beta\sigma$ -- a better parametrisation in relation to the standard superstratum in our three-dimensional notation, and plot the contour $C=0$ in the $(\alpha,\, \beta)$ plane. This is done in Fig.~\ref{fig:alpha_beta_plane}, using once again $q_1 = 5$ and $q_2 = 1$. We recover a similar result to the one presented in figure 12 of \cite{Ganchev:2021pgs}.

\vspace{0.5mm}
\vspace{1mm}

\vspace{0.5mm}
\vspace{1mm}

Because of \eqref{link_C_a}, the $C=0$ locus can also be identified with the ``$a\to 0$ limit'' of superstrata geometries. In the standard construction of the latter, there are two ways of taking that limit. To see that, one needs to transform back to the original $r$ coordinate, \eqref{xidef}. Then, one possibility is to keep $r$ finite and take $a\rightarrow0$, this leads to an asymptotically AdS$_3$ geometry with an infinitely-deep throat at the origin and a metric approaching that of extremal BTZ. The other option is to take both $r\to0$ and $a\rightarrow0$, while keeping the ratio $r/a$ fixed. This is a choice we have implicitly made by using the compactified radial variable, $\xi$. As we have seen, it produces an asymptotically AdS$_2\times S^1$ spacetime, as the AdS$_3$ region is pushed to an infinite distance, with a smooth AdS$_3$ cap at the origin.

The significance of the $a\rightarrow0$ limit for the standard superstratum and a resolution to the issue of whether a black hole solution can be obtained as a limit of such a horizonless geometry is discussed in \cite{Bena:2022sge}.

\begin{figure}[ht!]
	\centering
	\includegraphics[width=.5\textwidth]{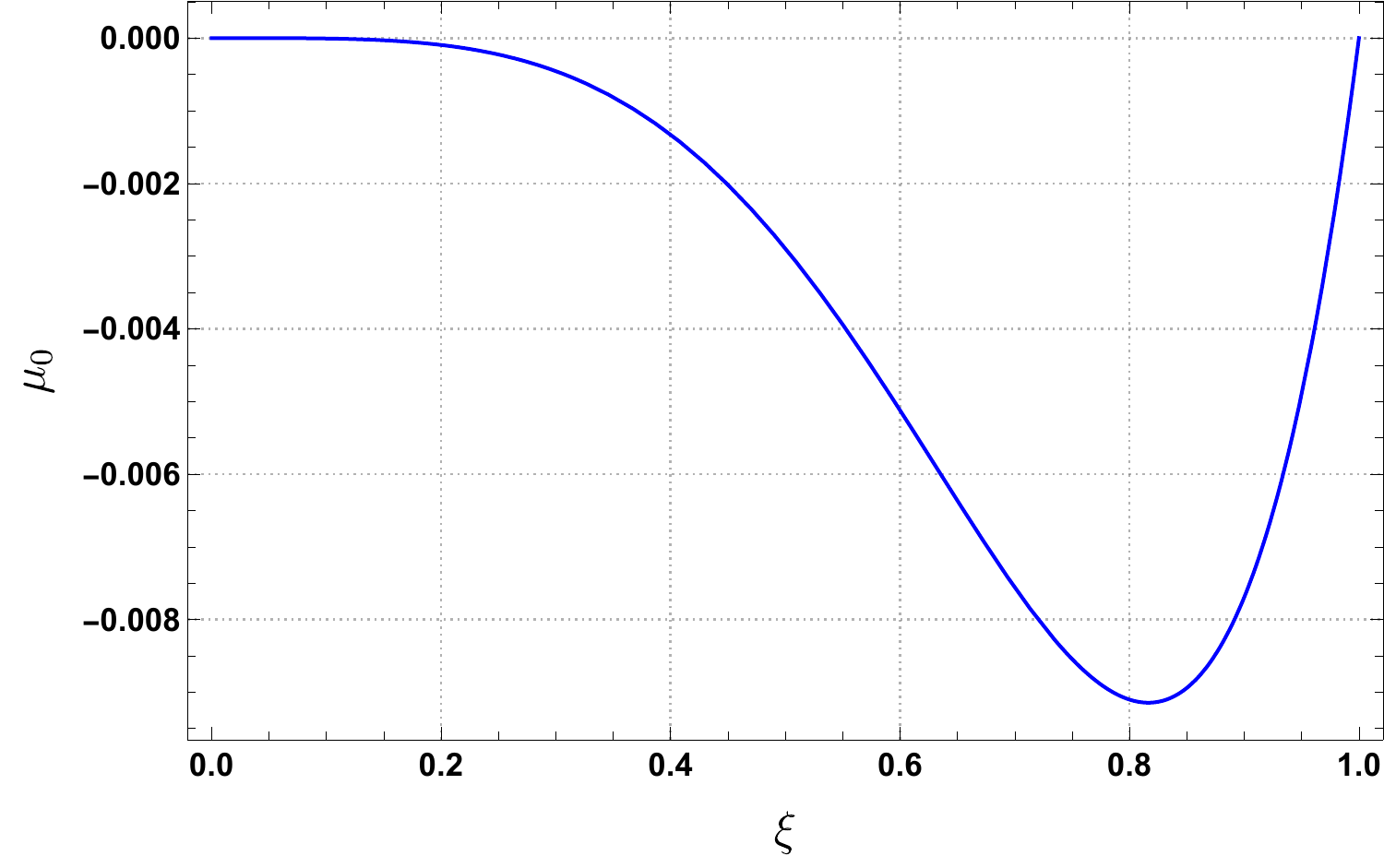}\hspace*{-1mm}
	\includegraphics[width=.5\textwidth]{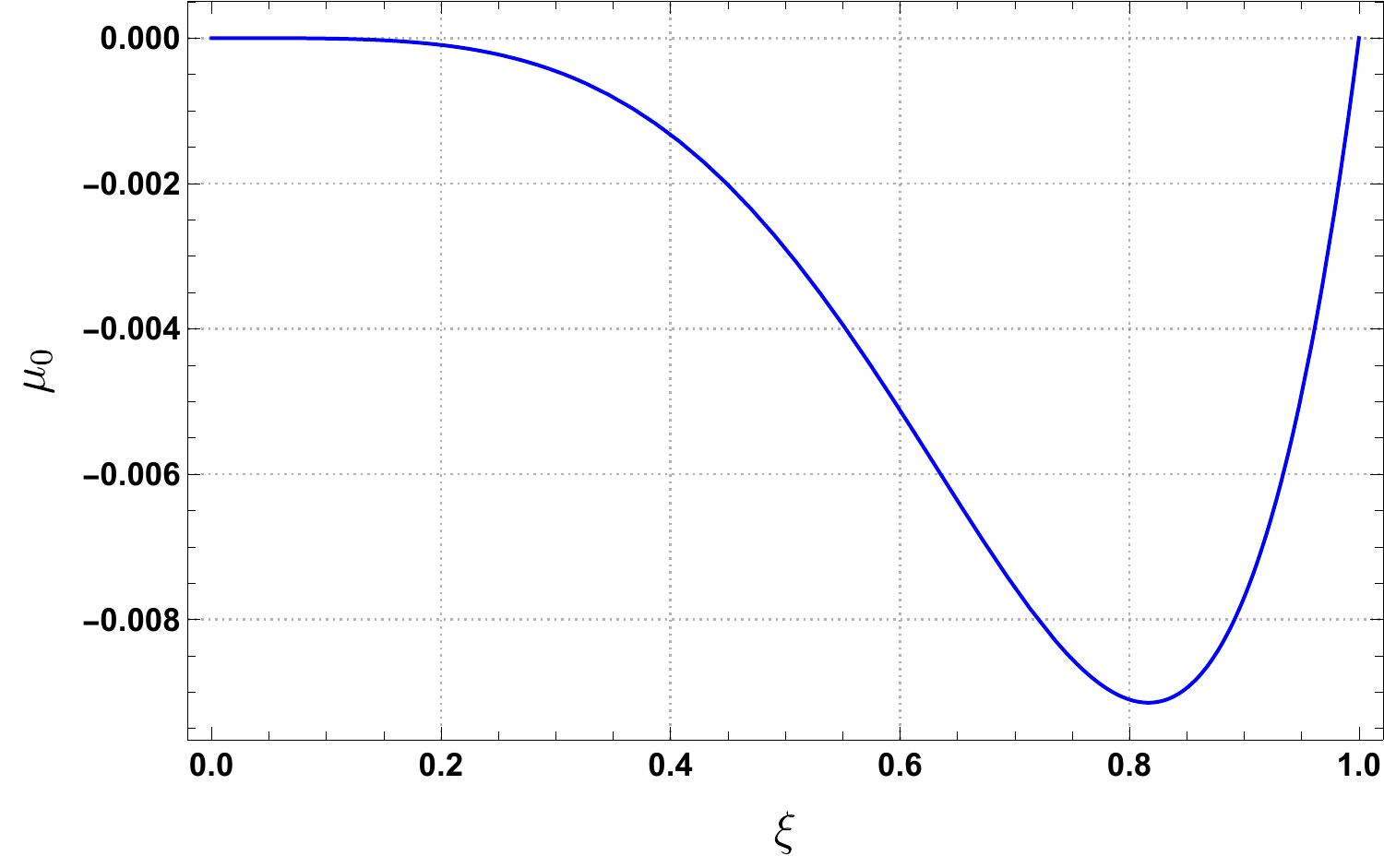}
	\caption{\it Plots of $\mu_0$ against $\xi$. In the first graph we have taken $\sigma =2$, $\beta = -\frac{q_1}{(2q_1 - 1)^2} + \frac{1}{10^6}$, $q_1=5$. The second graph corresponds to $\sigma =2$, $\beta = -\frac{q_1}{(2q_1 - 1)^2}$, $q_1=5$.}
	\label{fig:mu012}
\end{figure}

We have therefore shown, and depicted in Fig.~\ref{fig:gpsipsi12}, that our new superstrata, including the family in Section~\ref{sec:Example} that can be S-dualised to a solution with only NS-fields, can exhibit the deep scaling behavior of the standard superstrata as the momentum charge is increased towards its maximum value. Intriguingly enough, one can see from  (\ref{mu0res2}) and (\ref{Layer1bsol4}) that the scalars $\mu_0$,  $\mu_1$ and $\mu_2$ remain smooth and bounded even in the deepest such superstrata\footnote{Here, we are only considering solutions where the $\mu_j$ scalars are taken to vanish at infinity.}, except when $q_1 = q_2$. See, for example, Fig.~\ref{fig:mu012}, which shows a plot of $\mu_0$ on the special locus, \eqref{reduction2}, where $\mu_0=\mu_1$ and $\mu_2=0$. We use the same values of the parameters as in Fig.~\ref{fig:gpsipsi12}, and while that leads to different spacetime asymptotics, $\mu_0$ barely changes. This means that the elliptical deformation also remains bounded and goes nowhere near ``flattening the ellipse.''

\vspace{0.5mm}
\vspace{1mm}
\section{Final comments}
\label{sec:Conclusions}

Despite the very significant restrictions imposed by a consistent truncation to three-dimensional gauged supergravity, it is evident that the truncated theory still contains very interesting new branches of the moduli space of superstrata.  

Perturbative analysis around supertubes,  AdS$_3$ $\times S^3$ and within the dual CFT reveal the complete family of superstratum deformations \cite{deBoer:1998kjm,Ceplak:2018pws,Martinec:2017ztd,Martinec:2019wzw,Martinec:2020gkv}, however the challenge is  to ``integrate'' these perturbations to finite explorations of the moduli space.   If the perturbations lie within the linear BPS structure, then such integration is trivial and such deformations have been extensively exploited in the construction of superstrata.  Finite metric deformations are probably the most non-trivial moduli spaces because they involve tensorial quantities that explore the fully non-linear part of the BPS equations.   It is therefore very fortunate that  three-dimensional supergravity enables us to construct examples of precisely such geometries.

In this paper we have used  three-dimensional supergravity to construct superstrata based on  elliptically-deformed supertubes and uplift the results to six dimensions.  The resulting geometries have reduced symmetry (compared to standard superstrata) and it would have been very challenging to construct them directly in six dimensions.  However, now that the structure has been elucidated it may well be possible to generalize the results presented here to obtain new families of multi-mode superstrata directly in six dimensions.

There are also two important by-products of this work.

On the more formal side, we have obtained new classes of ambi-polar, hyper-K\"ahler manifolds in four dimensions:   geometries that only have a $U(1)$ isometry and this isometry is {\it not} tri-holomorphic.  One can think of these solutions as elliptical deformations of the usual two-centered Gibbons-Hawking geometries: the parameter that causes the elliptical deformation of the supertube appears in the hyper-K\"ahler geometry, and if the deformation is set to zero, one recovers a two-centered Gibbons-Hawking geometry with geometric charges $(n+1)$ and $-n$.  While it may be a computational challenge, one could, in principle, construct whole new families of six-dimensional superstrata based of this new deformed  hyper-K\"ahler manifold.

Perhaps the most important by-product is the construction of superstrata that lie  purely in the NS sector of the underlying string theory.    Standard superstrata involve momentum carriers encoded in the axion and the NS-fluxes, and these source further excitations in the dilaton and the RR fluxes \cite{Bena:2015bea}.  The careful pairing of these excitations is known as ``coiffuring,'' and this was essential to the construction of smooth solutions.    However, by opening up the metric deformations we have found a new way to create smooth solutions.  One can start with only the dilaton and the RR fluxes ($C_2$ and $C_6$)  of the original superstratum and smoothness can then be achieved through elliptical deformations of the base geometry.   These are the solutions described in Section \ref{sec:Example}.  Since one only has RR-fluxes and no axion, the S-dual solution contains only excitations of the metric, the NS flux and the dilation, and is therefore contained purely within the NS sector of the supergravity, or string. While the complexity of these new geometries might be a little daunting, we expect they might find invaluable applications in probing quantum microstructure via exact world-sheet methods  \cite{Martinec:2017ztd,Martinec:2019wzw,Martinec:2020gkv}.

Indeed, world-sheet techniques have been extensively studied for elliptically deformed supertubes in \cite{Martinec:2020gkv}, and we expect that such two-charge backgrounds correspond to our  (zero-momentum) superstrata with $q_1 = q_2 =1$. A particularly interesting limit of the supertubes considered in \cite{Martinec:2020gkv} arises when the ellipse is flattened out, and branes stretching across the supertube become massless.  Our results suggest that adding momentum  charge prevents the flattening of the supertube.  However,  branes  stretching across the supertube can still become massless because, while such branes have finite world-volume, their mass is hugely red-shifted in a very deep throat. This is very reminiscent of the ``W-branes'' of \cite{Martinec:2015pfa}, in which such wrapped branes retain a finite size in the fully back-reacted geometry but become massless, due to red-shifts, in the limit in which the throat of the microstate geometry becomes infinitely deep.

There are also a number of interesting questions within the purely three-dimensional setting. First there are some technical questions about the solutions presented here. We have focussed on solutions with $0 < \gamma_1, \gamma_2 < 1$, and especially on those with  $\gamma_2  = 1$.  While the $\gamma_i$ are required to have the same sign, we have not considered $\gamma_1, \gamma_2 \leq 0$.  More broadly, we have only looked at a very restricted Ansatz, defined in Section \ref{sub:qball}, which only includes the $(1,0,n)$ superstratum.  One could extend the Ansatz to include more general $(1,m,n)$ superstrata, and more general interactions between them. This may well lead to richer classes of six-dimensional geometries.  One could also explore multi-mode solutions even within the $(1,0,n)$ sector.

Then there are  non-BPS microstrata \cite{Ganchev:2021pgs,Ganchev:2021ewa} that can be obtained from supersymmetry-breaking excitations of our new classes of superstrata.  There are a huge range of possibilities here and some of them are being explored in forthcoming work \cite{GGHRW}, which considers a more general Ansatz in the spirit discussed in the previous paragraph, and offers new BPS geometries as well.   Indeed, in constructing the three-dimensional gauged supergravities that contain superstrata \cite{Mayerson:2020tcl}, the original goal was to use this formulation to obtain interesting, non-BPS microstrata.  As we remarked before, the three-dimensional formulation retains a remarkably rich BPS structure  that already has the capacity to inform and advance our understanding of far more general classes of superstrata and  advance exact world-sheet analysis of these backgrounds.   It seems that the non-BPS structure in three dimensions affords even more options for probing black-hole microstructure.

\section*{Acknowledgments}
\vspace{-2mm}
We would like to thank E.~Martinec and D.~Turton for valuable discussions.
The work of NPW is supported in part by the DOE grant DE-SC0011687. The work of BG, AH and NPW is supported in part by the ERC Grant 787320 - QBH Structure.


\appendix

\section{General solutions for the rest of the fields}
\label{app:restSols}

Here we explicitly present the general solutions to the fields that remain undetermined at the end of Section \ref{ss:layer1sol}. These can be obtained from \eqref{Omega1constraint},  \eqref{Phipots}, \eqref{numu0-2} and \eqref{BaseSystem} with the already derived solutions for $\mu_0$, $\mu_1$, $\mu_2$, $k$ and $\Omega_1$. 
\begin{gather}
\nu=\sqrt{\frac{\sigma}{1-\xi ^2}\Bigg(1-\frac{\big(\xi^{q_2}(1-\gamma_1\,\xi^{2\,q_1})+\beta\,\xi^{q_1}(1-\gamma_2\,\xi^{2\,q_2})\big)^2}{\big(\xi^{q_2}(1-\gamma_1\,\xi^{2\,q_1})-\beta \,\xi^{q_1}(1-\gamma_2\,\xi ^{2\,q_2})\big)^2}\Bigg)},
\end{gather}
\begin{gather}
\Phi_1=\frac{1}{2}\bigg(\frac{c_5}{1-\gamma_1\,\xi^{2\,q_1}}-\frac{c_4}{1-\gamma_2\,\xi ^{2\,q_2}}+\frac{(1+\gamma_1\,\xi^{2\,q_1})}{2(1-\gamma_1\,\xi^{2\,q_1})}\big((c_5-2)q_1\,\log\xi+c_7\big)\notag\\
-\frac{(1+\gamma_2\,\xi^{2\,q_2})}{2(1-\gamma_2\,\xi\,^{2\,q_2})}\big((c_4+2)q_2\,\log\xi+c_6\big)-\frac{\sigma}{2}\bigg)^{-1},
\end{gather}
\begin{gather}
\Phi_2=\frac{1}{2}\bigg(1-\bigg[\frac{\sigma\big(\xi^{q_2}(1-\gamma_1\,\xi^{2\,q_1})+\beta\,\xi^{q_1}(1-\gamma_2\,\xi^{2\,q_2})\big)}{2\big(\xi^{q_2}(1-\gamma_1\xi^{2\,q_1})-\beta\,\xi^{q_1}(1-\gamma_2\,\xi^{2\,q_2})\big)}-\frac{c_5}{1-\gamma_1\,\xi^{2\,q_1}}-\frac{c_4}{1-\gamma_2\,\xi ^{2\,q_2}}\notag\\
-\frac{(1+\gamma_1\,\xi^{2\,q_1})}{2(1-\gamma_1\,\xi^{2\,q_1})}\big((c_5-2)q_1\,\log\xi+c_7\big)-\frac{(1+\gamma_2\,\xi^{2\,q_2})}{2(1-\gamma_2\,\xi\,^{2\,q_2})}\big((c_4+2)q_2\,\log\xi+c_6\big)\bigg]^{-1}\bigg),
\end{gather}
\begin{gather}
\Psi_1=\frac{1}{4}\Big(q_2-\frac{2\,q_2}{1-\gamma_2\,\xi^{2\,q_2}}-q_1-\frac{2\,q_1}{1-\gamma_1\,\xi^{2\,q_1}}\Big)-\bigg[\big((c_4+2)\,q_2\,\log\xi+c_4+c_6\big)\frac{q_2\,\gamma_2\,\xi^{2\,q_2}}{(1-\gamma_2\,\xi^{2\,q_2})^2}\notag\\
+\frac{1}{2}\frac{c_4\,q_2}{1-\gamma_2\,\xi^{2\,q_2}}+\big((c_5-2)\,q_1\,\log\xi+c_5+c_7\big)\frac{q_1\,\gamma_1\,\xi^{2\,q_1}}{(1-\gamma_1\,\xi^{2\,q_1})^2}+\frac{1}{2}\,\frac{c_5\,q_1}{1-\gamma_1\,\xi^{2\,q_1}}+c_8\bigg]\times\notag\\
\times\frac{1}{2}\bigg(\frac{c_5}{1-\gamma_1\,\xi^{2\,q_1}}-\frac{c_4}{1-\gamma_2\,\xi ^{2\,q_2}}+\frac{(1+\gamma_1\,\xi^{2\,q_1})}{2(1-\gamma_1\,\xi^{2\,q_1})}\big((c_5-2)q_1\,\log\xi+c_7\big)\notag\\
-\frac{(1+\gamma_2\,\xi^{2\,q_2})}{2(1-\gamma_2\,\xi\,^{2\,q_2})}\big((c_4+2)q_2\,\log\xi+c_6\big)-\frac{\sigma}{2}\bigg)^{-1},
\end{gather}
\begin{gather}
\Psi_2=-\frac{1}{2}+\frac{q_1+q_2}{4}+\frac{q_1}{4}\Big[\frac{2\,\xi^{2\,q_2}(1-\gamma_1^2\,\xi^{4\,q_1})}{\xi^{2\,q_2}(1-\gamma_1\,\xi^{2\,q_1})^2-\beta^2\,\xi^{2\,q_1}(1-\gamma_2\,\xi^{2\,q_2})^2}+\frac{2}{1-\gamma_1\,\xi^{2\,q_1}}-4\Big]\notag\\
+\frac{q_2}{4}\Big[\frac{2}{1-\gamma_2\,\xi^{2\,q_2}}-\frac{2\,\beta^2\,\xi^{2\,q_1} (1-\gamma_2^2\,\xi^{4\,q_2})}{\xi^{2\,q_2}(1-\gamma_1\,\xi^{2\,q_1})^2-\beta^2\xi^{2\,q_1}(1-\gamma_2\,\xi^{2\,q_2})^2}-4\Big]\notag\\
+\frac{\gamma_2\,\xi^{2\,q_2}(2\,\gamma_1(q_1-q_2)(q_1+q_2)^2\xi^{2\,q_1}+\gamma_1^2\,q_2^2(2\,q_1+q_2)\,\xi^{4\,q_1}+q_2^3)-\gamma_1\,q_1^2\,\xi^{2\,q_1}(q_1+\gamma_2^2\,\xi^{4\,q_2}(q_1+2\,q_2))}{2\,\gamma_2\,q_2^2\,\xi^{2\,q_2}(1-\gamma_1\,\xi^{2\,q_1})^2-2\,\gamma _1\,q_1^2\xi^{2\,q_1}(1-\gamma_2\,\xi^{2\,q_2})^2}\notag\\
+\frac{1}{2}\bigg[\big((c_4+2)\,q_2\,\log\xi+c_4+c_6\big)\frac{q_2\,\gamma_2\,\xi^{2\,q_2}}{(1-\gamma_2\,\xi^{2\,q_2})^2}+\frac{1}{2}\frac{c_4\,q_2}{1-\gamma_2\,\xi^{2\,q_2}}+\frac{1}{2}\,\frac{c_5\,q_1}{1-\gamma_1\,\xi^{2\,q_1}}+c_8\notag\\
+\big((c_5-2)\,q_1\,\log\xi+c_5+c_7\big)\frac{q_1\,\gamma_1\,\xi^{2\,q_1}}{(1-\gamma_1\,\xi^{2\,q_1})^2}\bigg]\times\bigg[\frac{\sigma\big(\xi^{q_2}(1-\gamma_1\,\xi^{2\,q_1})+\beta\,\xi^{q_1}(1-\gamma_2\,\xi^{2\,q_2})\big)}{2\big(\xi^{q_2}(1-\gamma_1\xi^{2\,q_1})-\beta\,\xi^{q_1}(1-\gamma_2\,\xi^{2\,q_2})\big)}-\frac{c_5}{1-\gamma_1\,\xi^{2\,q_1}}\notag\\
-\frac{c_4}{1-\gamma_2\,\xi ^{2\,q_2}}-\frac{(1+\gamma_1\,\xi^{2\,q_1})}{2(1-\gamma_1\,\xi^{2\,q_1})}\big((c_5-2)q_1\,\log\xi+c_7\big)-\frac{(1+\gamma_2\,\xi^{2\,q_2})}{2(1-\gamma_2\,\xi\,^{2\,q_2})}\big((c_4+2)q_2\,\log\xi+c_6\big)\bigg]^{-1},
\end{gather}
\begin{gather}
\Omega_0=\Big[\frac{(1-\xi^2)^2 \big(\gamma_2\,q_2^2\,\xi^{2\,q_2}(1-\gamma_1\xi^{2\,q_1})^2-\gamma_1\,q_1^2\,\xi^{2\,q_1}(1-\gamma_2\,\xi^{2\,q_2})^2\big)}{4\,\xi^2(1-\gamma_1\,\xi^{2\,q_1})^2(1-\gamma_2\,\xi^{2\,q_2})^2}\times\notag\\
\times\Big(\frac{2\,c_5}{1-\gamma_1\,\xi^{2\,q_1}}-\frac{2\,c_4}{1-\gamma_2\,\xi^{2\,q_2}}-\sigma+\frac{(1+\gamma _1\,\xi ^{2 q_1})}{1-\gamma_1\,\xi^{2\,q_1}}\big((c_5-2)q_1\,\log\xi+c_7\big)\notag\\
-\frac{(1+\gamma_2\,\xi^{2\,q_2})}{1-\gamma_2\,\xi^{2\,q_2}}\big((c_4+2)q_2\,\log\xi+c_6\big)\Big)\times\Big(-\frac{2\,c_5}{1-\gamma_1\,\xi^{2\,q_1}}-\frac{2\,c_4}{1-\gamma_2\,\xi^{2\,q_2}}\notag\\
-\frac{(1+\gamma _1\,\xi ^{2 q_1})}{1-\gamma_1\,\xi^{2\,q_1}}\big((c_5-2)q_1\,\log\xi+c_7\big)-
\frac{(1+\gamma_2\,\xi^{2\,q_2})}{1-\gamma_2\,\xi^{2\,q_2}}\big((c_4+2)q_2\,\log\xi+c_6\big)\notag\\
+\frac{\sigma\big(\xi^{q_2}(1-\gamma_1\,\xi^{2\,q_1})+\beta\,\xi^{q_1}(1-\gamma_2\,\xi^{2\,q_2})\big)}{\xi^{q_2}(1-\gamma_1\,\xi^{2\,q_1})-\beta\,\xi^{q_1}(1-\gamma_2\,\xi^{2\,q_2})}\Big)\Big]^{1/2}.
\end{gather}


\begin{adjustwidth}{-1mm}{-1mm} 

\bibliographystyle{utphys}

\bibliography{microstates}       

\providecommand{\href}[2]{#2}\begingroup\raggedright\begin{thebibliography}{10}

\bibitem{Bena:2022ldq}
I.~Bena, E.~J. Martinec, S.~D. Mathur, and N.~P. Warner, ``{Snowmass White
  Paper: Micro- and Macro-Structure of Black Holes},''
  \href{http://arxiv.org/abs/2203.04981}{{\ttfamily arXiv:2203.04981
  [hep-th]}}.

\bibitem{Bena:2022rna}
I.~Bena, E.~J. Martinec, S.~D. Mathur, and N.~P. Warner, ``{Fuzzballs and
  Microstate Geometries: Black-Hole Structure in String Theory},''
  \href{http://arxiv.org/abs/2204.13113}{{\ttfamily arXiv:2204.13113
  [hep-th]}}.

\bibitem{Warner:1983du}
N.~P. Warner, ``{Some Properties of the Scalar Potential in Gauged Supergravity
  Theories},'' \href{http://dx.doi.org/10.1016/0550-3213(84)90286-4}{{\em Nucl.
  Phys. B} {\bfseries 231} (1984) 250--268}.

\bibitem{Godazgar:2014eza}
H.~Godazgar, M.~Godazgar, O.~Kr\"uger, H.~Nicolai, and K.~Pilch, ``{An
  SO(3)$\times$SO(3) invariant solution of $D=11$ supergravity},''
  \href{http://dx.doi.org/10.1007/JHEP01(2015)056}{{\em JHEP} {\bfseries 01}
  (2015) 056}, \href{http://arxiv.org/abs/1410.5090}{{\ttfamily arXiv:1410.5090
  [hep-th]}}.

\bibitem{Cvetic:2000dm}
M.~Cvetic, H.~Lu, and C.~Pope, ``{Consistent Kaluza-Klein sphere reductions},''
  \href{http://dx.doi.org/10.1103/PhysRevD.62.064028}{{\em Phys. Rev. D}
  {\bfseries 62} (2000) 064028},
  \href{http://arxiv.org/abs/hep-th/0003286}{{\ttfamily arXiv:hep-th/0003286}}.

\bibitem{Cvetic:2000zu}
M.~Cvetic, H.~Lu, and C.~Pope, ``{Consistent sphere reductions and universality
  of the Coulomb branch in the domain wall / QFT correspondence},''
  \href{http://dx.doi.org/10.1016/S0550-3213(00)00462-4}{{\em Nucl. Phys. B}
  {\bfseries 590} (2000) 213--232},
  \href{http://arxiv.org/abs/hep-th/0004201}{{\ttfamily arXiv:hep-th/0004201}}.

\bibitem{Nicolai:2001ac}
H.~Nicolai and H.~Samtleben, ``{N=8 matter coupled AdS(3) supergravities},''
  \href{http://dx.doi.org/10.1016/S0370-2693(01)00779-1}{{\em Phys. Lett.}
  {\bfseries B514} (2001) 165--172},
\href{http://arxiv.org/abs/hep-th/0106153}{{\ttfamily arXiv:hep-th/0106153
  [hep-th]}}.

\bibitem{Nicolai:2003bp}
H.~Nicolai and H.~Samtleben, ``{Chern-Simons versus Yang-Mills gaugings in
  three-dimensions},''
  \href{http://dx.doi.org/10.1016/S0550-3213(03)00569-8}{{\em Nucl. Phys.}
  {\bfseries B668} (2003) 167--178},
\href{http://arxiv.org/abs/hep-th/0303213}{{\ttfamily arXiv:hep-th/0303213
  [hep-th]}}.

\bibitem{Nicolai:2003ux}
H.~Nicolai and H.~Samtleben, ``{Kaluza-Klein supergravity on AdS(3) x S**3},''
  \href{http://dx.doi.org/10.1088/1126-6708/2003/09/036}{{\em JHEP} {\bfseries
  09} (2003) 036},
\href{http://arxiv.org/abs/hep-th/0306202}{{\ttfamily arXiv:hep-th/0306202
  [hep-th]}}.

\bibitem{Deger:2014ofa}
N.~S. Deger, H.~Samtleben, O.~Sar{\i}o\u{g}lu, and D.~Van~den Bleeken, ``{A
  supersymmetric reduction on the three-sphere},''
  \href{http://dx.doi.org/10.1016/j.nuclphysb.2014.11.014}{{\em Nucl. Phys.}
  {\bfseries B890} (2014) 350--362},
\href{http://arxiv.org/abs/1410.7168}{{\ttfamily arXiv:1410.7168 [hep-th]}}.

\bibitem{Samtleben:2019zrh}
H.~Samtleben and O.~Sar{\i}o\u{g}lu, ``{Consistent $S^3$ reductions of
  six-dimensional supergravity},''
  \href{http://dx.doi.org/10.1103/PhysRevD.100.086002}{{\em Phys. Rev.}
  {\bfseries D100} no.~8, (2019) 086002},
\href{http://arxiv.org/abs/1907.08413}{{\ttfamily arXiv:1907.08413 [hep-th]}}.

\bibitem{Mayerson:2020tcl}
D.~R. Mayerson, R.~A. Walker, and N.~P. Warner, ``{Microstate Geometries from
  Gauged Supergravity in Three Dimensions},''
  \href{http://dx.doi.org/10.1007/JHEP10(2020)030}{{\em JHEP} {\bfseries 10}
  (2020) 030}, \href{http://arxiv.org/abs/2004.13031}{{\ttfamily
  arXiv:2004.13031 [hep-th]}}.

\bibitem{Bena:2015bea}
I.~Bena, S.~Giusto, R.~Russo, M.~Shigemori, and N.~P. Warner, ``{Habemus
  Superstratum! A constructive proof of the existence of superstrata},''
  \href{http://dx.doi.org/10.1007/JHEP05(2015)110}{{\em JHEP} {\bfseries 05}
  (2015) 110},
\href{http://arxiv.org/abs/1503.01463}{{\ttfamily arXiv:1503.01463 [hep-th]}}.

\bibitem{Bena:2016ypk}
I.~Bena, S.~Giusto, E.~J. Martinec, R.~Russo, M.~Shigemori, D.~Turton, and
  N.~P. Warner, ``{Smooth horizonless geometries deep inside the black-hole
  regime},'' \href{http://dx.doi.org/10.1103/PhysRevLett.117.201601}{{\em Phys.
  Rev. Lett.} {\bfseries 117} no.~20, (2016) 201601},
\href{http://arxiv.org/abs/1607.03908}{{\ttfamily arXiv:1607.03908 [hep-th]}}.

\bibitem{Tian:2016ucg}
W.~Tian, ``{Multicenter superstrata},''
  \href{http://dx.doi.org/10.1103/PhysRevD.94.066011}{{\em Phys. Rev. D}
  {\bfseries 94} no.~6, (2016) 066011},
  \href{http://arxiv.org/abs/1607.08884}{{\ttfamily arXiv:1607.08884
  [hep-th]}}.

\bibitem{Bena:2016agb}
I.~Bena, E.~Martinec, D.~Turton, and N.~P. Warner, ``{Momentum Fractionation on
  Superstrata},'' \href{http://dx.doi.org/10.1007/JHEP05(2016)064}{{\em JHEP}
  {\bfseries 05} (2016) 064},
\href{http://arxiv.org/abs/1601.05805}{{\ttfamily arXiv:1601.05805 [hep-th]}}.

\bibitem{Bena:2017xbt}
I.~Bena, S.~Giusto, E.~J. Martinec, R.~Russo, M.~Shigemori, D.~Turton, and
  N.~P. Warner, ``{Asymptotically-flat supergravity solutions deep inside the
  black-hole regime},'' \href{http://dx.doi.org/10.1007/JHEP02(2018)014}{{\em
  JHEP} {\bfseries 02} (2018) 014},
\href{http://arxiv.org/abs/1711.10474}{{\ttfamily arXiv:1711.10474 [hep-th]}}.

\bibitem{Heidmann:2019zws}
P.~Heidmann and N.~P. Warner, ``{Superstratum Symbiosis},''
  \href{http://dx.doi.org/10.1007/JHEP09(2019)059}{{\em JHEP} {\bfseries 09}
  (2019) 059}, \href{http://arxiv.org/abs/1903.07631}{{\ttfamily
  arXiv:1903.07631 [hep-th]}}.

\bibitem{Heidmann:2019xrd}
P.~Heidmann, D.~R. Mayerson, R.~Walker, and N.~P. Warner, ``{Holomorphic Waves
  of Black Hole Microstructure},''
  \href{http://dx.doi.org/10.1007/JHEP02(2020)192}{{\em JHEP} {\bfseries 02}
  (2020) 192}, \href{http://arxiv.org/abs/1910.10714}{{\ttfamily
  arXiv:1910.10714 [hep-th]}}.

\bibitem{Shigemori:2020yuo}
M.~Shigemori, ``{Superstrata},''
  \href{http://dx.doi.org/10.1007/s10714-020-02698-8}{{\em Gen. Rel. Grav.}
  {\bfseries 52} no.~5, (2020) 51},
  \href{http://arxiv.org/abs/2002.01592}{{\ttfamily arXiv:2002.01592
  [hep-th]}}.

\bibitem{Ganchev:2021pgs}
B.~Ganchev, A.~Houppe, and N.~P. Warner, ``{Q-balls meet fuzzballs: non-BPS
  microstate geometries},''
  \href{http://dx.doi.org/10.1007/JHEP11(2021)028}{{\em JHEP} {\bfseries 11}
  (2021) 028}, \href{http://arxiv.org/abs/2107.09677}{{\ttfamily
  arXiv:2107.09677 [hep-th]}}.

\bibitem{Ganchev:2021ewa}
B.~Ganchev, S.~Giusto, A.~Houppe, and R.~Russo, ``{$\hbox {AdS}_3$ holography
  for non-BPS geometries},''
  \href{http://dx.doi.org/10.1140/epjc/s10052-022-10133-2}{{\em Eur. Phys. J.
  C} {\bfseries 82} no.~3, (2022) 217},
  \href{http://arxiv.org/abs/2112.03287}{{\ttfamily arXiv:2112.03287
  [hep-th]}}.

\bibitem{Houppe:2020oqp}
A.~Houppe and N.~P. Warner, ``{Supersymmetry and superstrata in three
  dimensions},'' \href{http://dx.doi.org/10.1007/JHEP08(2021)133}{{\em JHEP}
  {\bfseries 08} (2021) 133}, \href{http://arxiv.org/abs/2012.07850}{{\ttfamily
  arXiv:2012.07850 [hep-th]}}.

\bibitem{Ganchev:2021iwy}
B.~Ganchev, A.~Houppe, and N.~P. Warner, ``{New superstrata from
  three-dimensional supergravity},''
  \href{http://dx.doi.org/10.1007/JHEP04(2022)065}{{\em JHEP} {\bfseries 04}
  (2022) 065}, \href{http://arxiv.org/abs/2110.02961}{{\ttfamily
  arXiv:2110.02961 [hep-th]}}.

\bibitem{Ceplak:2018pws}
N.~{\v C}eplak, R.~Russo, and M.~Shigemori, ``{Supercharging Superstrata},''
  \href{http://dx.doi.org/10.1007/JHEP03(2019)095}{{\em JHEP} {\bfseries 03}
  (2019) 095}, \href{http://arxiv.org/abs/1812.08761}{{\ttfamily
  arXiv:1812.08761 [hep-th]}}.

\bibitem{Bena:2011dd}
I.~Bena, S.~Giusto, M.~Shigemori, and N.~P. Warner, ``{Supersymmetric Solutions
  in Six Dimensions: A Linear Structure},''
  \href{http://dx.doi.org/10.1007/JHEP03(2012)084}{{\em JHEP} {\bfseries 1203}
  (2012) 084},
\href{http://arxiv.org/abs/1110.2781}{{\ttfamily arXiv:1110.2781 [hep-th]}}.

\bibitem{Giusto:2013rxa}
S.~Giusto, L.~Martucci, M.~Petrini, and R.~Russo, ``{6D microstate geometries
  from 10D structures},''
  \href{http://dx.doi.org/10.1016/j.nuclphysb.2013.08.018}{{\em Nucl.Phys.}
  {\bfseries B876} (2013) 509--555},
\href{http://arxiv.org/abs/1306.1745}{{\ttfamily arXiv:1306.1745 [hep-th]}}.

\bibitem{Martinec:2017ztd}
E.~J. Martinec and S.~Massai, ``{String Theory of Supertubes},''
  \href{http://dx.doi.org/10.1007/JHEP07(2018)163}{{\em JHEP} {\bfseries 07}
  (2018) 163}, \href{http://arxiv.org/abs/1705.10844}{{\ttfamily
  arXiv:1705.10844 [hep-th]}}.

\bibitem{Martinec:2019wzw}
E.~J. Martinec, S.~Massai, and D.~Turton, ``{Little Strings, Long Strings, and
  Fuzzballs},'' \href{http://dx.doi.org/10.1007/JHEP11(2019)019}{{\em JHEP}
  {\bfseries 11} (2019) 019},
\href{http://arxiv.org/abs/1906.11473}{{\ttfamily arXiv:1906.11473 [hep-th]}}.

\bibitem{Martinec:2020gkv}
E.~J. Martinec, S.~Massai, and D.~Turton, ``{Stringy Structure at the BPS
  Bound},'' \href{http://dx.doi.org/10.1007/JHEP12(2020)135}{{\em JHEP}
  {\bfseries 12} (2020) 135}, \href{http://arxiv.org/abs/2005.12344}{{\ttfamily
  arXiv:2005.12344 [hep-th]}}.

\bibitem{Bena:2004de}
I.~Bena and N.~P. Warner, ``{One ring to rule them all ... and in the darkness
  bind them?},'' {\em Adv. Theor. Math. Phys.} {\bfseries 9} (2005) 667--701,
\href{http://arxiv.org/abs/hep-th/0408106}{{\ttfamily arXiv:hep-th/0408106}}.

\bibitem{Gibbons:1978tef}
G.~W. Gibbons and S.~W. Hawking, ``{Gravitational Multi - Instantons},''
  \href{http://dx.doi.org/10.1016/0370-2693(78)90478-1}{{\em Phys. Lett. B}
  {\bfseries 78} (1978) 430}.

\bibitem{Balasubramanian:2000rt}
V.~Balasubramanian, J.~de~Boer, E.~Keski-Vakkuri, and S.~F. Ross,
  ``{Supersymmetric conical defects: Towards a string theoretic description of
  black hole formation},''
  \href{http://dx.doi.org/10.1103/PhysRevD.64.064011}{{\em Phys. Rev.}
  {\bfseries D64} (2001) 064011},
\href{http://arxiv.org/abs/hep-th/0011217}{{\ttfamily arXiv:hep-th/0011217}}.

\bibitem{Maldacena:2000dr}
J.~M. Maldacena and L.~Maoz, ``{De-singularization by rotation},'' {\em JHEP}
  {\bfseries 12} (2002) 055,
\href{http://arxiv.org/abs/hep-th/0012025}{{\ttfamily arXiv:hep-th/0012025}}.

\bibitem{Mathur:2001pz}
S.~D. Mathur, ``{Gravity on AdS(3) and flat connections in the boundary CFT},''
  \href{http://arxiv.org/abs/hep-th/0101118}{{\ttfamily arXiv:hep-th/0101118}}.

\bibitem{Lunin:2004uu}
O.~Lunin, ``{Adding momentum to D1-D5 system},''
  \href{http://dx.doi.org/10.1088/1126-6708/2004/04/054}{{\em JHEP} {\bfseries
  04} (2004) 054},
\href{http://arxiv.org/abs/hep-th/0404006}{{\ttfamily arXiv:hep-th/0404006}}.

\bibitem{Bena:2008wt}
I.~Bena, N.~Bobev, and N.~P. Warner, ``{Spectral Flow, and the Spectrum of
  Multi-Center Solutions},''
  \href{http://dx.doi.org/10.1103/PhysRevD.77.125025}{{\em Phys. Rev.}
  {\bfseries D77} (2008) 125025},
\href{http://arxiv.org/abs/0803.1203}{{\ttfamily arXiv:0803.1203 [hep-th]}}.

\bibitem{Bena:2022sge}
I.~Bena, N.~Ceplak, S.~Hampton, Y.~Li, D.~Toulikas, and N.~P. Warner,
  ``{Resolving Black-Hole Microstructure with New Momentum Carriers},''
  \href{http://arxiv.org/abs/2202.08844}{{\ttfamily arXiv:2202.08844
  [hep-th]}}.

\bibitem{deBoer:1998kjm}
J.~de~Boer, ``{Six-dimensional supergravity on S**3 x AdS(3) and 2-D conformal
  field theory},'' \href{http://dx.doi.org/10.1016/S0550-3213(99)00160-1}{{\em
  Nucl. Phys.} {\bfseries B548} (1999) 139--166},
\href{http://arxiv.org/abs/hep-th/9806104}{{\ttfamily arXiv:hep-th/9806104
  [hep-th]}}.

\bibitem{Martinec:2015pfa}
E.~J. Martinec and B.~E. Niehoff, ``{Hair-brane Ideas on the Horizon},''
  \href{http://dx.doi.org/10.1007/JHEP11(2015)195}{{\em JHEP} {\bfseries 11}
  (2015) 195},
\href{http://arxiv.org/abs/1509.00044}{{\ttfamily arXiv:1509.00044 [hep-th]}}.

\bibitem{GGHRW}
B.~Ganchev, S.~Giusto, A.~Houppe, R.~Russo, and N.~P. Warner $\!\!$, {\it to
  appear}.

\end{thebibliography}\endgroup

\end{adjustwidth}

\end{document}